\documentclass[pre,singlecolumn,noshowpacs,amsmath,amssymb,nofootinbib]{revtex4}

\usepackage{mathrsfs}
\usepackage{graphicx}
\usepackage{dcolumn}
\usepackage{bm}
\usepackage{color}
\usepackage{times}
\usepackage{ulem}
\usepackage{footnote}

\newcommand{\tb}{\textbf}

\newcommand{\ra}{\rangle}
\newcommand{\la}{\langle}

\newcommand{\eps}{\epsilon}

\newcommand{\bmGamma}{\bm{\Gamma}}
\newcommand{\mclS}{\mathcal{S}}
\newcommand{\TR}{{\rm TR}}
\newcommand{\ST}{{\rm ST}}

\newcommand{\RT}{{\rm RT}}
\newcommand{\unbiased}{{\rm unbiased}}
\newcommand{\naive}{{\rm NLLS}}
\newcommand{\smooth}{{\rm smooth}}
\newcommand{\sample}{{\rm sample}}
\newcommand{\threshold}{{\rm th}}
\newcommand{\stat}{{\rm stat}}
\newcommand{\temp}{{\rm temp}}

\newcommand{\DFA}{{\rm DFA}}

\newcommand{\LMF}{{\rm LMF}}

\newcommand{\RLS}{{\rm RLS}}
\newcommand{\OLS}{{\rm OLS}}

\newcommand{\acf}{{(a)}}
\newcommand{\psd}{{(s)}}
\newcommand{\spectra}{{(s)}}

\DeclareMathOperator*{\argmin}{arg\,min}

\begin{document}
\title{Quantitative statistical analysis of order-splitting behaviour of individual trading accounts \\in the Japanese stock market over nine years}

\author{Yuki Sato}
\affiliation{
	Department of Physics, Graduate School of Science, Kyoto University, Kyoto 606-8502, Japan
}
\author{Kiyoshi Kanazawa}
\email{kiyoshi@scphys.kyoto-u.ac.jp}
\affiliation{
	Department of Physics, Graduate School of Science, Kyoto University, Kyoto 606-8502, Japan
}	
\date{\today}

\begin{abstract}
	Econophysics aims to understand the macroscopic behaviour of financial markets from the underlying microscopic decision-making dynamics. In particular, the order splitting of large metaorders is one of the most important trading strategies in this literature: while traders have large potential metaorders, they split the large orders into small pieces (called child orders) to minimise market impact. This strategic behaviour is believed to be important because it is a promising candidate for the microscopic origin of the long-range correlation (LRC) in the persistent order flow. Indeed, in 2005, Lillo, Mike, and Farmer (LMF) introduced a simple microscopic model of the order-splitting traders to predict the asymptotic behaviour of the LRC from the microscopic dynamics, even quantitatively. The plausibility of this scenario has been investigated by T\'oth et al. 2015 at a qualitative level. However, no solid support has been presented yet on the quantitative prediction by the LMF model in the lack of large microscopic datasets. In this report, we have provided the first quantitative statistical analysis of the order-splitting behaviour at the level of each trading account. We analyse a large dataset of the Tokyo stock exchange (TSE) market over nine years, including the account data of traders (called {\it virtual servers}). The virtual server is a unit of trading accounts in the TSE market, and we can effectively define the trader IDs by an appropriate preprocessing. We apply a strategy clustering to individual traders in terms of market orders to identify the order-splitting traders and the random traders. The length distribution of metaorders are empirically estimated for each stock every year. For most of the stocks, we find that the metaorder length distribution obeys power laws with exponent $\alpha$, such that $P(L)\propto L^{-\alpha-1}$ with the metaorder length $L$, as theoretically assumed in the LMF model. By analysing the sign correlation of order flow $C(\tau)\propto \tau^{-\gamma}$, we draw the scatterplot between $\alpha$ and $\gamma$, directly confirming the LMF prediction $\gamma \approx \alpha-1$. Furthermore, we discuss how to estimate the total number of the splitting traders only from public data via the ACF prefactor formula in the LMF model. Our work provides the first quantitative evidence of the LMF model, strongly supporting the order-splitting hypothesis as the origin of LRC.
\end{abstract}

\maketitle

\section{Introduction.}
	The ultimate goal of statistical physics is to reveal the macroscopic behaviours of physical systems from their microscopic dynamics, and physicists have broadly applied this concept to interdisciplinary topics, such as financial markets, beyond traditional physics~\cite{EconPhys_Stanley, EconPhys_Slanina, BouchaudText, PhysRep2022}. Recently, econophysicists have greatly benefitted from high-frequency financial data on the microscopic level of individual traders~\cite{KanazawaPRL, KanazawaPRE, Sueshige2018, Sueshige2019}. In this report, we focus on the microscopic origin of the long-range correlation (LRC) of the order flow by providing the first systematic statistical analysis of a large comprehensive dataset on the level of individual trading accounts.

	In recent financial markets, traders are required to submit limit orders or market orders for their trading activities. The limit order is an option to show the traders' potential will to buy or sell the stock by specifying their prices in advance. All the limit orders are collected as the limit-order book, which displays the current potential prices for transactions. Limit-order submissions are called {\it liquidity provision} in the economic context and highly appreciated because they stabilise the market. On the other hand, if traders wish to transact immediately, they can submit market orders to buy or sell the stock at the best price (i.e., the highest bid or lowest ask price). Market-order submissions are called {\it liquidity consumption}, in contrast to limit-order submissions. In other words, financial markets are composed of the flows of limit orders and market orders, and the main target of this report is the market-order flow, particularly, regarding its persistence.

	The market-order flow exhibits strong persistence in financial markets: the buy (sell) market orders tend to follow another buy (sell) market order for a long time. In other words, once you observe a buy (sell) order, it is more likely to observe buy (sell) orders in future (e.g., a typical order-sign series is given by $\{\eps(t)\}_t=\{+1,+1,-1,+1,+1,+1, \dots\}$, where $\eps(t) = +1$ ($\eps(t) = -1$) denotes a buy (sell) market order). More quantitatively, the power-law decay of the sign autocorrelation function (ACF) characterises this phenomenon~\cite{BouchaudText,Bouchaud2003,Lillo2004, BouchaudReview, Farmer2004}, called the long-range correlation (LRC) in this report:
  \begin{equation}
    C(\tau) := \lim_{t\to \infty}\la \eps(t)\eps(t+\tau)\ra \approx \frac{c_0}{\tau^{\gamma}}, \>\>\> \gamma \in (0,1)
  \end{equation}
  for a large timelag $\tau \gg 1$ with the characteristic power-law exponent $\gamma$ and the prefactor $c_0$. Here $\la A \ra$ denotes the ensemble average of the stochastic variable $A$. This LRC is ubiquitously observed in various financial markets, such as stocks~\cite{Biais, Bouchaud2003, Bouchaud2006, Lillo2004, Eisler}, foreign exchange (FX)~\cite{FX}, crypto-currency markets~\cite{Donier}, and, therefore, is believed to be an essential of the financial market microstructure. 

	\begin{figure*}
		\centering
		\includegraphics[width=135mm]{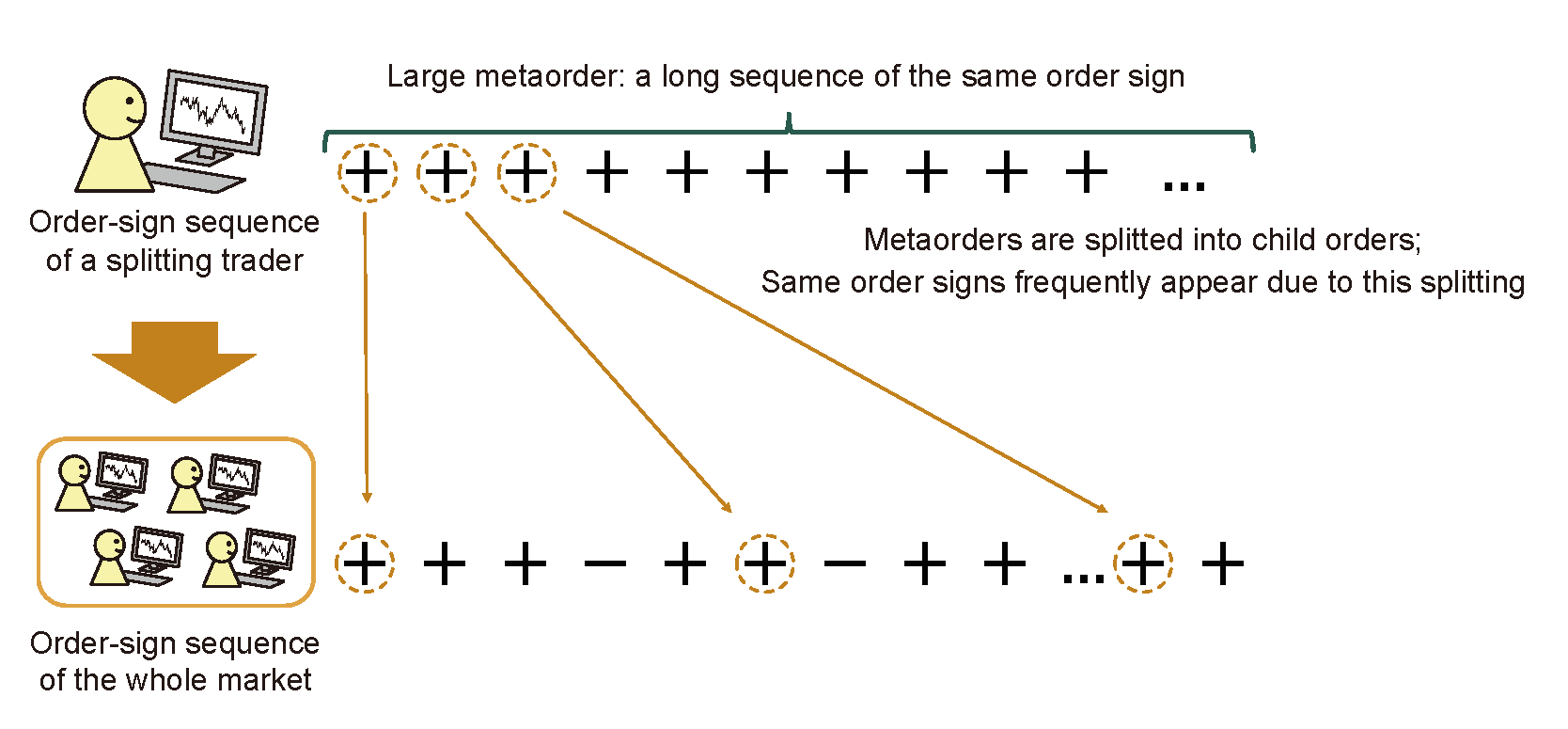}
		\caption{
			Schematic of the order-splitting hypothesis. In this hypothesis, several traders hold large latent orders (called metaorders) and split them into small child orders. The plus (minus) sign ``$+$" (``$-$") represents a buy (sell) market order. By definition, the child orders have the same order sign; thus, the order-sign sequence of the whole market exhibits a long memory due to this order splitting. 
		}
		\label{fig:OrderSplittingHypothesis}
	\end{figure*}
		Then, what is the microscopic origin of the LRC? One of the most promising hypotheses is the order-splitting behaviour at the level of individual traders~\cite{BouchaudText} (see Fig.~\ref{fig:OrderSplittingHypothesis} as a schematic). According to this hypothesis, several traders hold large latent orders (called the {\it metaorders}). Typically, the size of such a metaorder is much larger than the revealed liquidity on the order book, and, therefore, the traders have no choice but to split the metaorder into a long sequence of small market orders (called the {\it child orders}) to minimise the transaction cost (called the {\it market impact}), naturally leading to the LRC in the sign ACF. From the side of empirical analyses, this scenario has been supported qualitatively. While it is difficult to perform statistical analyses of comprehensive data, including all trader IDs, various fragmented data support the plausibility of the order-splitting hypothesis~\cite{BouchaudText}. In addition, Ref.~\cite{Toth2015} provided a crucial evidence on the qualitative importance of the order-splitting based on a comprehensive dataset: the authors of Ref.~\cite{Toth2015} decomposed the ACF $C(\tau)$ into the contribution by the same traders $C_{\mathrm{same}}(\tau)$ and that by other traders $C_{\mathrm{other}}(\tau)$. They finally showed that the former contribution is much larger than the latter one as $|C_{\mathrm{other}}(\tau)/C_{\mathrm{same}}(\tau)| \ll 1$ for large $\tau$, suggesting the strong relevance of the order-splitting behaviours at least qualitatively

	To organize this scenario more quantitatively and precisely, Lillo, Mike, and Farmer proposed a simple theoretical microscopic model (called the LMF model~\cite{LMF}) of the order-splitting behaviour at the level of individual traders in 2005. They have provided a clear explanation of the macroscopic LRC nature from the microscopic dynamics. Specifically, they assume that the length $L$ of metaorders obeys the power law distribution 
	\begin{equation}
		\label{eq:powerLaw_rho}
		P(L)\approx L^{-\alpha-1}, \>\>\> \alpha > 1. 
	\end{equation}
	Under this assumption, they made a powerful quantitative prediction that the macroscopic behaviour of the LRC should be directly related to the microscopic parameter of the model, such that 
	\begin{equation}
		\gamma = \alpha - 1. 
		\label{eq:LMF_prediction}
	\end{equation}

	While the LMF model has been regarded as one of the stylised microscopic models of order-splitting for 18 years, its empirical foundation has not been fully verified in particular for its quantitative prediction~\eqref{eq:LMF_prediction}. While it is obviously appealing to provide its direct empirical verification, several severe difficulties have prohibited such empirical research: (i)~ Estimating the microscopic parameter $\alpha$ requires special comprehensive datasets, including all trader IDs. However, such datasets are scarce from the viewpoint of data availability. (ii)~The quantitative confirmation of the prediction~\eqref{eq:LMF_prediction} is expected to require very large datasets. Indeed, because $\gamma$ empirically distributes between zero to one, the estimation errors in the power-law exponents $\alpha$ and $\gamma$ should be controlled roughly less than $0.1$ even for drawing the scatterplot. This fact suggests that larger datasets are necessary than usual financial data analyses. (iii)~Furthermore, the intrinsic long-memory character of the LRC essentially causes the slower convergence of its statistical estimator in estimating $\gamma$ than usual (in our estimation, at least an order-sign sequence longer than 0.5 million transactions is necessary for obtaining even one datapoint of $\gamma$). Due to these three fundamental problems, the direct verification of the LMF model has been a crucial unsolved problem in econophysics. 

	In this report, together with the companion Letter~\cite{SatoPRLCompanion}, we present the first quantitative verification of the LMF prediction~\eqref{eq:LMF_prediction} by analysing a large high-frequency dataset on the level of trading accounts. We have studied a large comprehensive order-book detaset on the Tokyo Stock exchange (TSE) market, the biggest stock-exchange platform in Japan. This dataset covers the nine-years period from 2012 to 2020 for all the stocks. Remarkably, this dataset includes the {\it virtual server ID}, which is a unit of the trader accounts in the TSE platform. By appropriately analysing the virtual server IDs, this data allows us to virtually track the trading behaviour of all individual traders. Based on this dataset, firstly, this report addresses the trading-strategy classification on the level of individual traders in terms of market orders. We classify all traders into the random traders (RTs) and the splitting traders (STs) by the binomial test, directly confirming the presense of the STs for most of the stocks. We next measure the metaorder-length (run-length) distribution among the STs for each stock. As assumed in the LMF model, we confirm that the metaorder length for the STs obeys power-laws, such that $P(L)\propto L^{-\alpha-1}$ for large $L$ with $\alpha>1$. By measuring the power-law exponent $\gamma$ in the LRC ($C(\tau)\propto \tau^{-\gamma}$), we provide the scatterplot between $\alpha$ and $\gamma$ and then directly verify the LMF prediction~\eqref{eq:LMF_prediction} even at the quantitative level. As the last discussion, we study the estimation of the total number of the order-splitting traders from public data via the LMF theory regarding the prefactor $c_0$.

	This report is organised as follows. We describe our dataset, the TSE market rule, and our mathematical notation in Sec.~\ref{sec:DataDescription}. In Sec.~\ref{sec:literature_review}, we provide a short review on the LMF model and its related literature. In Sec.~\ref{sec:Measure_MetaOrder_alpha}, we apply our strategy-clustering algorithm to measure $\alpha$ for each stock. In Sec.~\ref{sec:measure_gamma}, we describe our statistical method to measure $\gamma$ for each stock. The scatterplot between $\alpha$ and $\gamma$ is provided in Sec.~\ref{sec:ScatterPlot} as the main result. We conclude our report in Sec.~\ref{sec:conclusion}. Eleven appendices follows to supplement the main text.

\section{Data description and the market rule}\label{sec:DataDescription}
	\subsection{Dataset}
		Here we describe our high-frequency dataset on the TSE market in detail. Our dataset was provided by the Japan Exchange (JPX) Group, Inc., which is the platform manager of the TSE market. This dataset covers all the stocks in the TSE market during the nine-years period, from the 4th January 2012 to the 30th December 2020. This comprehensive dataset includes the order ID (i.e., the unique identifier to track the life cyle of any order), type (i.e., buy or sell), order type (i.e., limit order, cancellation order, and market order), price, and virtual server ID. See also Appendix~\ref{sec:app:dataAvailability} for the data availability.

\begin{figure}
		\centering
		\includegraphics[width=150mm]{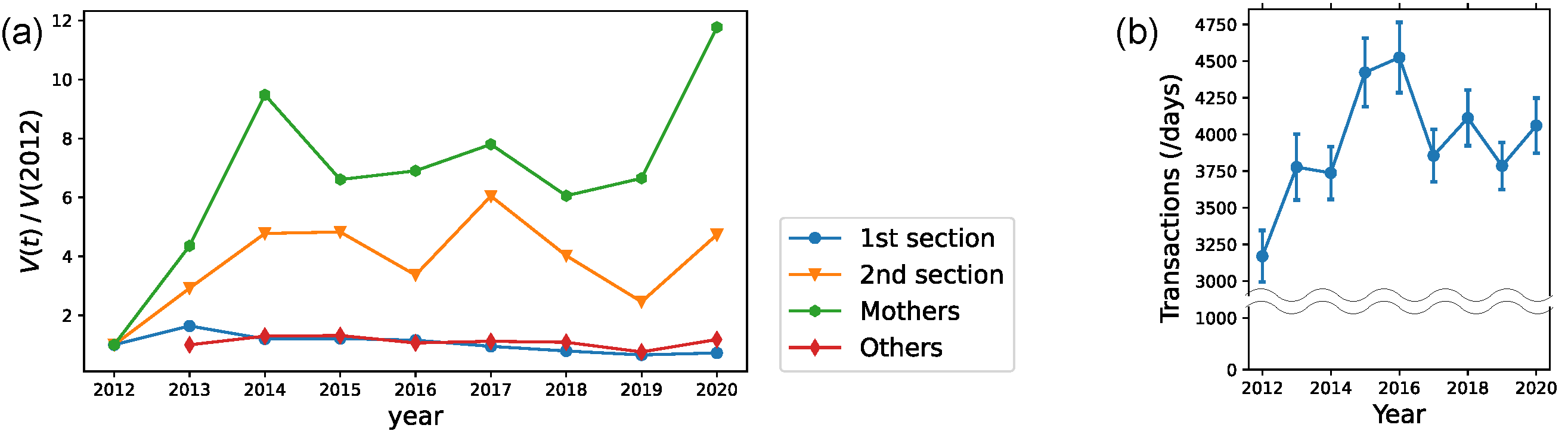}
		\caption{
			Summary statistics on the TSE markets from 2012 to 2020. 
			(a)~Daily transaction volume on the arrowhead system $V(t)$ normalised by the volume in 2012 (i.e. $V(2012)$). This figure is based on the public data provided by the TSE about the monthly/yearly transaction-volume statistics. This data shows the increase in transaction volumes, particularly in the second section and Mothers markets.
			(b)~Typical daily transaction numbers each year for the stock markets used for this study. As described later, we studied only stocks whose yearly transaction number is over 0.5 million. We calculated each stock's yearly average transaction number and then took its simple average across stocks for this plot. 
		}
		\label{fig:transactions_on_Arrowhead}
	\end{figure}

		The TSE trading system is called the {\it arrowhead}. The arrowhead system was updated three times in our dataset. For example, while the reaction speed of the arrowhead was $2$ millisecond at the beginning of our dataset, it was updated to be $1$ millisecond, $0.3$ millisecond, and $0.2$ millisecond on 17th July 2012, 24th September 2015, and 5th November 2019, respectively. The high reaction speed of the arrowhead facilitates tradings and the number of transactions increases in the TSE second section and Mothers (see Fig.~\ref{fig:transactions_on_Arrowhead}(a)). Because sufficient observations are crucial for precise measurement of the power-law exponent $\gamma$, it is expected that the measurement precision of $\gamma$ will be better as time goes by, particularly after 2015 (see Fig.~\ref{fig:transactions_on_Arrowhead}(b)).
	\subsection{Definition of trader IDs: virtual server IDs and trading desks}
		\begin{table}
			\begin{tabular}{c|c|c|c}
				Order ID & Virtual server ID & Type         & Trading desk ID \\ \hline \hline
				O1       & V1                & submission   & T1              \\
				O1       & V1                & cancellation & T1              \\
				O2       & \tb{V1}     			 & submission   & T1              \\
				O2       & \tb{V2}					 & cancellation & T1             
			\end{tabular}
			\caption{
				Schematic idea of the {\it trading desk} ID. Let us consider the case where a trader issues submission and cancellation orders. Typically, the virtual server IDs between these orders are identical (see the order flow of the order ID ``O1'' with the same virtual server ID ``V1''). At the same time, if the trader possesses two virtual servers ``V1'' and ``V2'', the trader can issue the cancellation order from a different virtual server ``V2''. For such a case, we infer that both virtual servers ``V1'' and ``V2'' are issued by the same trader and then allocate a single trading desk ID ``T1''. In this report, the trading desk ID is regarded as representating an effective membership and is called the {\it trader ID} for short. 
			}
			\label{table:tradingdesks}
		\end{table}
		One of the remarkable advantages of our dataset is that it includes the virtual server ID. The virtual server is a unit of the trader accounts in the TSE. The virtual server ID is a consistent identifier of the TSE participants, but, technically, it is not completely equivalent to the membership ID. Indeed, there is an option for any trader to possess several virtual server IDs. For example, there is a limit on the number of submission from a single virtual server during a fixed time interval. Some traders possess several virtual servers to avoid this submission limit for high-frequency tradings.

		One of the technical solutions to this problem is to use the {\it trading desks} as an effective proxy of the membership ID, which was introduced by the work by Goshima, Tobe, and Uno~\cite{Goshima2019}. The outline of their idea is to aggregate several virtual server IDs to allocate a unified ID (i.e., the trading desk) if we detect that the virtual servers are associated with the same membership\footnote{
			While the virtual server IDs are kept identical for most of the periods, they were shuffled when the arrowhead system was updated on the 24th September 2015 and on the 4th November 2019.}. 
		For example, let us consider the case where a trader possesses two virtual servers ``V1" and ``V2", submits a new limit order, and then cancel it finally. Typically, the virtual server IDs are identical between the submission and cancellation orders (see the order flow in Table~\ref{table:tradingdesks} with the order ID ``O1''). Sometimes, however, there are non-typical cases where the virtual server IDs are not identical between the submission and cancellation orders (see the order flow in Table~\ref{table:tradingdesks} with the order ID ``O2''): e.g., when the trader submits a submission order with the order ID ``O2" from the virtual server ``V1" and subsequently submits a cancelleation order with the order ID ``O2" from the virtual server ``V2", it is reasonable to infer that both virtual server IDs ``V1" and ``V2" are associated to the identical trader. The concept of the trading desk is to merge these two server IDs to allocate a single label as the effective trader ID (i.e. ``T1'' in Table~\ref{table:tradingdesks}). For the detailed implementation, see Refs.~\cite{Goshima2019,Hirano2020}. In this report, the trading desk is regarded as the effective membership ID and is called the {\it trader ID} for short.
		
		It should be noted that in Japan, the TSE is not the sole stock market available. Various venues, including the proprietary trading system (PTS), exist where identical stocks can be traded. In addition, if a ``final client", e.g., a mutual fund, can trade with different market members, their multiple IDs might be aggregated. In such cases, our trader ID might combine multiple metaorders from diverse clients, which would then be treated as a unified metaorder in our analyses.

	\subsection{Market rule}
		Here we describe the market rule in the TSE market. The TSE provides three types of trading periods: (i)~the opening auctions (during 08:00-09:00 and 12:05-12:30), (ii)~the continuous double auctions (during 09:00-11:30 and 12:30-15:00), and (iii)~the closing auctions (at 11:30 and 15:00). Throughout this report, the time is based on the Japan Standard Time (JST, UTC+9).

		During the opening auctions, all the orders are collected but wait for their transaction until the fixed transaction time 9:00 or 12:30. During the continuous double auctions, all orders can be immediately executed under the time priority rule if the supply and demand match. In this report, we focus on the continuous double auction periods. 

		In TSE, there are three types of orders: the limit order, the cancellation order, and the market order. Any limit order is composed of the price, the volume, and the type (i.e., bid or ask). When a trader is potentially willing to buy (sell) the specified volume of the stock at the specified price, the trader will submit a bid (ask) limit order. The limit order can be cancelled if the trader is unwilling to buy (sell) the stock anymore.

	\subsection{Limit order book}
		While the background knowledge of the limit order book (LOB) is not essential in understanding our main findings, we briefly explain several important concepts related to the LOB, since they are useful in discussing the possible implications of our findings.

		All the live limit orders are collected to form the limit order book (LOB). A part of the LOB is publicly displayed and is used as an information source for decision making by traders. The most important part of the LOB is the best bid (ask) price, defined by the highest bid (lowest ask) price in the LOB. Also, the market spread, defined by the difference between the best ask and bid prices, is an important measure of the effective transaction cost. We note that submitting limit orders is regarded as beneficial contribution to the market liquidity. Indeed, if there is a plenty of bid and ask limit orders, anyone will be able to make a large volume of transaction with a small transaction cost. In this sense, traders keeping a plenty of bid and ask limit orders are sometimes called {\it liquidity providers} or {\it market makers}. 

		On the other hand, the market order is the order to make a transaction at the available best prices. For example, if a trader submits the buy (sell) market order, the trader immediately buys (sells) the stock at the best ask (bid) prices. In contrast to the limit-order submissions, submitting market orders are regarded as the liquidity consumption. Therefore, traders who submit market orders are sometimes called {\it liquidity consumers} or {\it takers}. 

	\subsection{Mathematical notation}
		\subsubsection{The fundamental quantities}
			\label{sec:math_notation}
			\begin{figure*}
				\includegraphics[width=180mm]{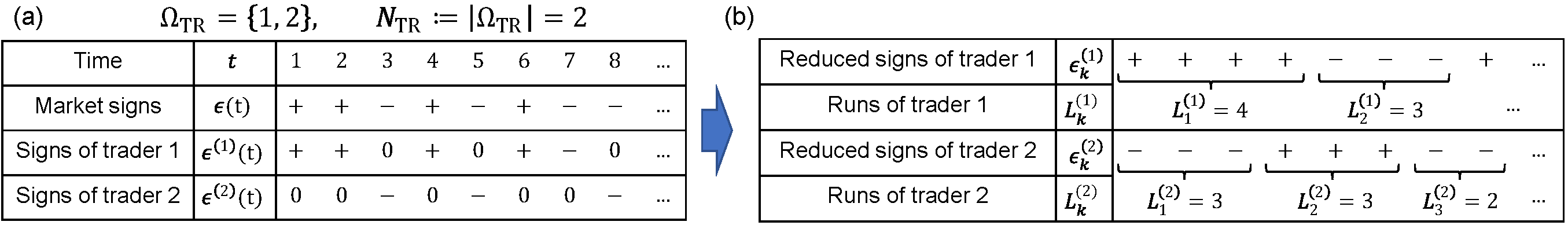}
				\caption{
					(a)~Schematic example of the fundamental quantities $\bmGamma: = \left(\Omega_{\TR}, \{\eps(t)\}_t, \{\eps^{(i)}(t)\}_{t,i}\right)$ for the case $N_{\TR}=|\Omega_{\TR}|=2$. Here, $+$ ($-$) is an abbreviation of $+1$ ($-1$), representing a buy (sell) order.
					(b)~Reduced-sign sequences $\{\eps_k^{(i)}\}_{k}$ are defined by removing zeros from the original order sequences $\{\eps^{(i)}(t)\}_{t}$ for $i\in \Omega_{\TR}$. Runs $\{L_k^{(i)}\}_k$ are also defined as the numbers of the successively same signs for the trader $i$.
				}
				\label{fig:MathNotation}
			\end{figure*}
			Here we explain the mathematical notation for our analyses of one datapoint. In this report, we focus on the following fundamental quantities (see Fig.~\ref{fig:MathNotation}(a) for a scheme): 
			\begin{itemize}
				\item $\Omega_{\TR}$: the set of all trader IDs. The total size of the traders is finite, such that $|\Omega_{\TR}|=N_{\TR} < \infty$. Therefore, the trader IDs can be rewritten as $\Omega_{\TR} = \{i \>|\> i=1,2,\dots, N_{\TR} \}$ without losing generality. 
				\item $\eps(t)$: the market-order sign at the discrete time $t \in \bm{N}$ in the whole market, with the set of natural integers $\bm{N}=\{1,2,\dots\}$. Here, the order sign $\eps(t)=+1$ ($\eps(t)=-1$) signifies the buy (sell) market order and the time $t$ is measured as a positive integer time (called {\it tick time}), incremented every transaction. The total number of the market orders is denoted by $N_{\eps} := |\{\eps(t)\}_t|$, which is finite for real data analyses.  
				\item $\eps^{(i)}(t)$: the market-order sign issued by the trader $i \in \Omega_{\TR}$ at time $t$. If the trader $i$ did not issue any order at time $t$, $\eps^{(i)}_t$ is set to be zero: $\eps^{(i)}(t)=0$. By definition, an identity holds such that 
				\begin{equation}
					\epsilon(t) = \sum_{i \in \Omega_{\TR}} \eps^{(i)}(t). \label{eq:RelationIndMacro}
				\end{equation}
			\end{itemize}
			Here, the fundamental set $\bmGamma:= \left(\Omega_{\TR}, \{\eps(t)\}_{t \in \bm{N}}, \{\eps^{(i)}(t)\}_{t\in \bm{N}, i\in \Omega_{\TR}}\right)$ completely characterises our analyses. We note that the volume information on any market order is not used in this report. 
		
		\subsubsection{Other important quantities}
			In addition, we can define the following quantities as derivatives of the fundamental quantities $\bmGamma$ (see Fig.~\ref{fig:MathNotation}(b) for a schematic): 
			\begin{itemize}
				\item $C(\tau)$: the market autocorrelation function (ACF) with the timelag $\tau \geq 0$, defined by $C(\tau):= \la \eps(t)\eps(t+\tau)\ra$, where $\la A\ra$ denotes the ensemble average of any stochastic quantity $A$. 
				\item $\{\eps^{(i)}_k\}_{k\in \bm{N}}$: the reduced order-sign sequences, by removing zeros from the original order-sign sequences $\{\eps^{(i)}(t)\}_{t\in \bm{N}}$. The total number of the market orders for the trader $i$ is denoted by $N^{(i)}_{\rm MO}:=|\{\eps^{(i)}_k\}_{k}|$, which can be finite. 

				\item $\{L^{(i)}_k\}_{k\in \bm{N}}$: the runs for the reduced-sign sequences $\{\eps^{(i)}_k\}_{k\in \bm{N}}$ for the trader $i \in \Omega_{\TR}$. For a given reduced-sign sequences $\{\eps^{(i)}_k\}_{k\in \bm{N}}$ of the trader $i \in \Omega_{\TR}$, we define the {\it runs} similarly to the Wald-Wolfowitz runs test~\cite{RunsTest}. In other words, for $\{\eps^{(i)}_k\}_k$, we count the numbers of adjacent equal elements (e.g., $L_1^{(1)}=4$ and $L_2^{(1)}=3$ for $\{++++---+...\}$) to define the runs $\{L^{(i)}_k\}_k$ (see Fig.~\ref{fig:MathNotation}). 
			\end{itemize}
			As will be explained in Sec.~\ref{sec:Measure_MetaOrder_alpha}, we apply a strategy clustering in terms of market orders to define the following classes of traders: 
			\begin{itemize}
				\item $\Omega_{\RT}$: the set of the random traders (RTs). 
				\item $\Omega_{\ST}$: the set of the splitting traders (STs). 
			\end{itemize}
			By definition, we have $\Omega_{\TR} = \Omega_{\RT} \cup \Omega_{\ST}$.

		\subsubsection{Sample label for the integrated statistical analysis}
			More technically, the fundamental quantities $\bmGamma$ are defined for each datapoint. By introducing a sample label $s \in \mclS$ to identify each datapoint with the sample set $\mclS$, $\{\bmGamma_s\}_{s\in \mclS}$ is finally analysed as the integrated statistical analysis to produce Figs.~\ref{fig:scatterplot_empirical} and \ref{fig:emp_estimate_N} (e.g., the scatterplot between $\alpha$ and $\gamma$ in Sec.~\ref{sec:ScatterPlot}). One datapoint $\bmGamma_s$ corresponds to an yearly order-sign sequence for one stock market (i.e., the label $s$ signifies the set of stock ticker code and year). The total number of the datapoints is denoted by $N_{\mclS}:= |\mclS|$. However, if the expression clearly makes sense in the context, the sample label $s$ is often omitted for brevity.

		\subsubsection{Filter on the sample markets}
			In this report, we focus on the markets whose total transaction number is over 0.5 million, such that $N_{\eps} > 5\times 10^5$. This filter is introduced to suppress the estimation errors in the power-law exponents $\alpha$ and $\gamma$. 

		\subsubsection{Other mathematical notation}
			We next describe our notation for the probability theory. The probability density function (PDF) characterises the probability that the stochastic variable $x'$ resides in the range $[x,x+dx)$ as $P(x)dx$. The complementary cumulative distribution function (CCDF) is defined by $P_>(x):=\int_x^{\infty} dyP(y)$. 

			For a given series $\{x_k\}_k$, we can define the empirical PDF and CCDF as  
			\begin{equation}
				P(x) := \frac{1}{\left|\{x_k\}_k\right|}\sum_{k} \delta(x-x_k), \>\>\> 
				P_>(x) := \int_{x}^\infty dyP(y) = \frac{N_>(x)}{\left|\{x_k\}_k\right|},
			\end{equation}
			where $N_>(x):= \int_x^\infty \sum_k \delta(y-x_k)dy$ is the total number of the elements larger than $x$, and $\delta(x)$ is the Dirac delta function. 

\subsection{Data preprocessing}
	Here we explain our data preprocessing to extract data by removing the influence of intraday seasonality. Intraday seasonality is one of the stylised facts in financial markets~\cite{BouchaudText}, and the market activity typically exhibits high intensity around the opening and closing times of the auctions (called the U-shape profile). Indeed, we confirmed the U-shape profile in terms of the market-activity statistics (see Appendix.~\ref{sec:app:IntradaySeasonality}).
	
	This intraday seasonality should be considered in interpreting the results of any data analysis because there are various factors unique to the opening and closing times of the auctions (such as the lifestyle of traders and the position management~\cite{HusseinPhD}, for example). Such factors are not included in the LMF model; therefore, the data during such high-activity periods are not suitable for the data calibrations. 
	
	For these reasons, we used the market-order sign sequence during the continuous double auction periods with the ten-minutes sequences excluded around the opening and closing auctions. In other words, we used the data from 9:10 to 11:20 and from 12:40 to 14:50 as a daily order-sign sequence. The daily order-sign sequences are segmented on the yearly basis for each stock to obtain one datapoint $\bmGamma$.

\section{Literature review on the LMF model}\label{sec:literature_review}
	This section reviews the LMF model in terms of the model setup, quantitative prediction, and the current qualitative empirical evidence. This section aims to provide background knowledge on this econophysics topic for the general audience to clarify the novelty of our results. Since this review section is prepared independently of the other sections, readers interested only in our main results may skip this section.

	\subsection{Microscopic model: the original LMF model}
		\begin{figure*}
			\centering
			\includegraphics[width=170mm]{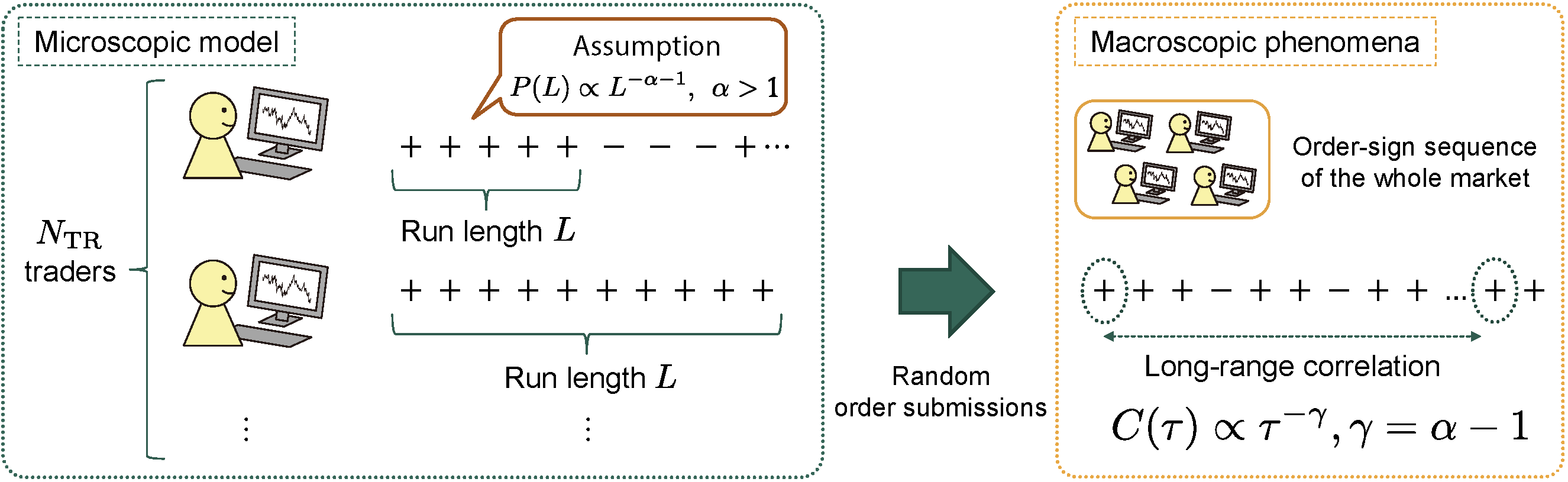}
			\caption{
				Schematic of the Lillo-Mike-Farmer (LMF) model proposed in Ref.~\cite{LMF}. At the microscopic dynamics, the total number of traders is $N_{\TR}$ and all the traders are assumed to be splitting traders (STs). STs hold large metaorders and they randomly split them into small orders. Here we assume that the run length $L$ obeys the power-law distribution $P(L)\propto L^{-\alpha-1}$ with $\alpha>1$. At the macroscopic dynamics, the order-sign sequence of the whole market exhibits the long-range correlation $C(\tau)\propto \tau^{-\gamma}$. Furthermore, the LMF model theoretically predicts $\gamma = \alpha - 1$ as Eq.~\eqref{eq:review_LMP_quantPred}, connecting the macroscopic power-law exponent $\gamma$ and the microscopic exponent $\alpha$.
				The state variables and model parameters are summarised in Table~\ref{table:summary_stateVariables_parameters}.
			}
			\label{fig:LMF_setup}
		\end{figure*}
		\begin{table}
			\centering
			\begin{tabular}{c|l||c|l}
			State variable    & Meaning                                  & Model parameters & Meaning                          \\ \hline \hline
			$\eps(t)$         & Order sign in the whole market           & $N_{\TR}$  & Total number of the traders      \\
			$\{\eps^{(i)}(t)\}_i$ & Order sign of the trader $i$             &  $N_{\eps}$ & Total number of the transactions     \\
			$\{R^{(i)}(t)\}_i$   & Remaining metaorder length of trader $i$ & $P(L)\propto L^{-\alpha-1}$, $\alpha>1$     & Metaorder length distribution
			\end{tabular}
			\caption{
				The summary of the state variables and the model parameters for the LMF model~\eqref{eq:def:LMF}. 
			}
			\label{table:summary_stateVariables_parameters}
		\end{table}
		Let us assume that the total number of traders $N_{\TR}>0$ is a time-constant positive integer and the volume of any market order is always the minimum executable unit for simplicity. For any trader $i \in \Omega_{\TR}$, two microscopic variables are defined: $\bm{z}^{(i)}(t):=(\eps^{(i)}(t), R^{(i)}(t))$, where $\eps^{(i)}$ is the order sign of the metaorder and $R^{(i)}$ is the remaining volume of the metaorder. The macroscopic variable of the market is given by the market-order sign $\eps(t)$. The LMF model is formulated as the Markovian stochastic process for the state variable $\bm{Z}:=(\eps; \bm{z}_1,...,\bm{z}_{N_{\TR}})$ on the discrete time $t \in \bm{N}$. 

		The concrete dynamics of this model is given by the following stochastic difference equations (SDEs, see Fig.~\ref{fig:LMF_setup}): 
		\begin{subequations}
			\label{eq:def:LMF}
			At the time $t+1$, a trader $i=\pi(t+1)$ is randomly selected with the uniform distribution, such that
			\begin{equation}
				P_{t+1}(\pi) = \frac{1}{N_{\TR}} \>\>\> \mbox{ for any $\pi \in \Omega_{\TR}$}.
			\end{equation}
			The $\pi(t+1)$-th trader executes their metaorder with the order sign: 
			\begin{equation}
				\eps (t+1) = \eps^{(\pi(t+1))}(t).
			\end{equation}
			After the execution by the trader $\pi$, the remaining volume $R^{(\pi)}(t+1)$ decreases by one if $R^{(\pi)}(t) > 1$. If all the metaorder is executed (i.e., $R^{(\pi)}(t)=1$), the metaorder and its sign are randomly reset for the trader $\pi$. In summary, the dynamics of $\bm{z}^{(i)}$ is given as follows for all $i \in \Omega_{\TR}$: 
			\begin{align}
				R^{(i)}(t+1) &= 
				\begin{cases}
					R^{(i)}(t) & \mbox{if $i \neq \pi(t+1)$} \\
					R^{(i)}(t) - 1 & \mbox{if $i = \pi(t+1)$ and $R^{(i)}(t)>1$} \\
					L & \mbox{if $i= \pi(t+1)$ and $R^{(i)}(t)=1$; $L$ obeys $P(L)$}
				\end{cases} \\
				\eps^{(i)}(t+1) &= 
				\begin{cases}
					\eps^{(i)}(t) & \mbox{if $i \neq \pi(t+1)$ or $R^{(i)}(t)>1$} \\
					+1 & \mbox{with prob. $1/2$, if $i = \pi(t+1)$ and $R^{(i)}(t)=1$} \\
					-1 & \mbox{with prob. $1/2$, if $i = \pi(t+1)$ and $R^{(i)}(t)=1$}
				\end{cases}
			\end{align}
			with an independent and identically distributed (IID) random integer number $L>0$ obeying the discrete PDF $P(L)$. 
		\end{subequations}
		The set of the SDEs \eqref{eq:def:LMF} completely characterises the $(2N_{\TR}+1)$-dimensional Markovian dynamics with the state variable $\bm{Z}(t)$ on the discrete time $t\in \bm{N}$. In this sense, the SDEs \eqref{eq:def:LMF} are the fundamental ``equations of motion" for the LMF model at the microscopic level of the financial dynamics. We will consider the dynamics of this stochastic process until the final time $t=N_{\eps}:=|\{\eps(t)\}_t|$ (i.e., the total number of the transactions). See Table~\ref{table:summary_stateVariables_parameters} for the summary of the state variables and the model parameters.

		In this framework, all traders are assumed to simply split their metaorders without complicated strategies according to the order-splitting hypothesis, and the discrete PDF $P(L)$ can be interpreted as the distribution of the metaorder lengths (or the run lengths). For the consistency with the realistic data analysis, it is a customary to assume the power-law metaorder distribution:
		\begin{equation*}
			P(L) \propto L^{-\alpha-1} \mbox{ for large }L
		\end{equation*}
		with a realistic value~\cite{BouchaudText,Vaglica,Bershova} around $\alpha\approx 1.5$ (see Appendix~\ref{app:random_integer_powerlaw} for the detailed implementation in generating power-law random numbers). We can straightforwardly generalise this model to introduce heterogeneity of splitting strategies (see Ref.~\cite{SatoJSP2023}).
		
		\subsection{Quantitative prediction: from micro to macro}
		\begin{figure*}
			\centering
			\includegraphics[width=70mm]{"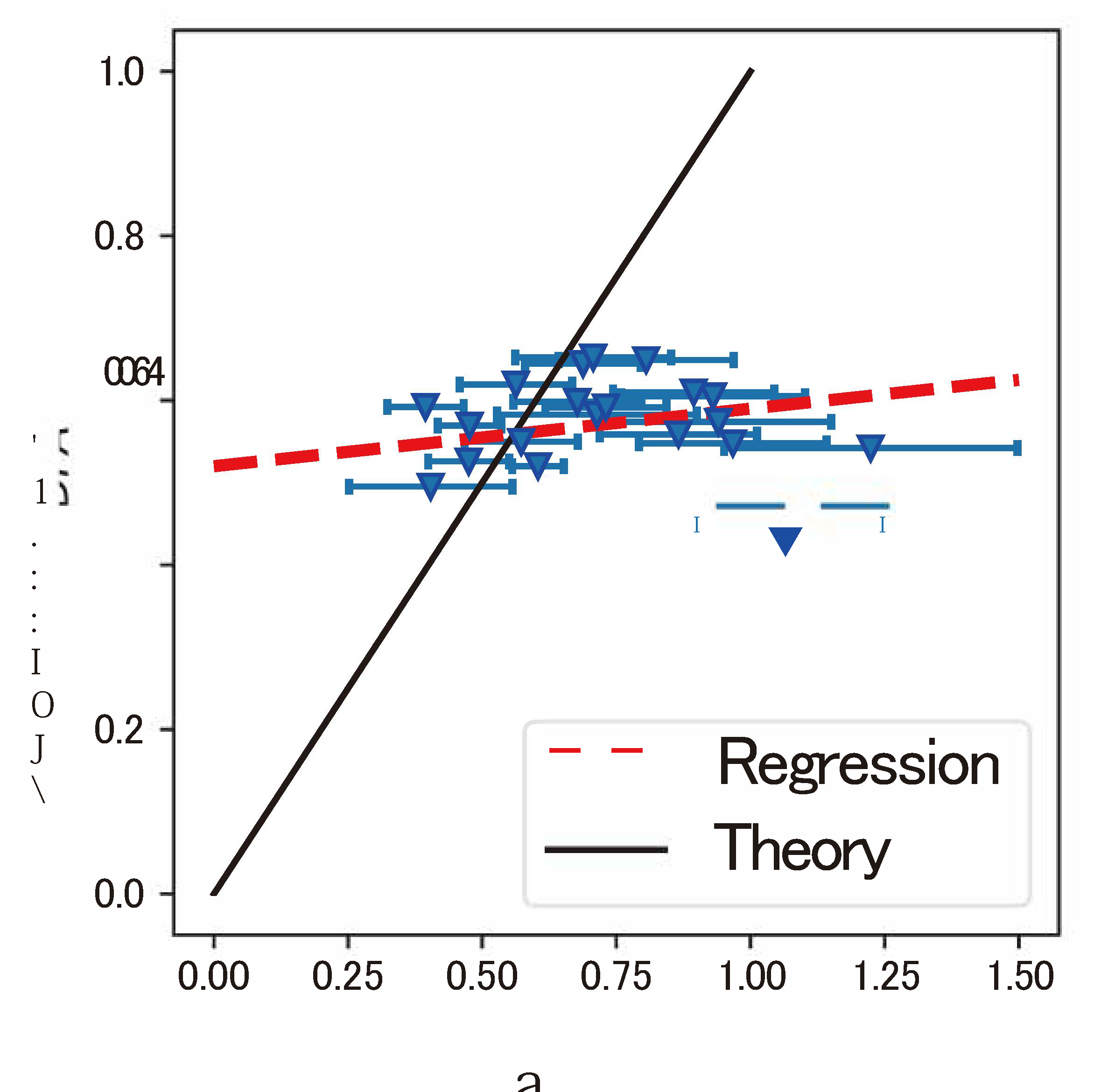"}
			\caption{
					Verification of the LMF prediction~\eqref{eq:LMF_prediction} in the previous work~\cite{LMF}. We extracted the data from the figure 7 of Ref.~\cite{LMF} by measuring the coodinates of the datapoints using Adobe Illustrator. Here we additionally plot the regression line (led) with the slope $0.07$ (statistically insignificant), far from the theoretical coefficient $1$. We note that they measured $\gamma$ based on the DFA (see the related discussion in Sec.~\ref{sec:measure_gamma}).
			}
			\label{fig:LMFPaper}
		\end{figure*}
		Since the microscopic model is fixed as the high-dimensional Markovian stochastic process~\eqref{eq:def:LMF}, the macroscopic character of this model can be deduced in principles. Such a statistical-mechanical program was provided by the original paper~\cite{LMF} by the Lillo, Mike, and Farmer. Indeed, the ACF of the order-sign sequence $\{\eps(t)\}_t$ is asymptotically given by 
		\begin{equation}
			C(\tau) := \lim_{t\to \infty} \left<\eps(t)\eps(t+\tau)\right> \propto 
			\tau^{-\gamma}, \>\>\> \gamma := \alpha - 1
			\label{eq:review_LMP_quantPred}
		\end{equation}
		for large $\tau$. This formula implies that the macroscopic parameter $\gamma$ is directly related to the microscopic parameter $\alpha$. In this report, the expression ``the quantitative prediction of the LMF model'' refers to this relationship~\eqref{eq:review_LMP_quantPred}. Note that this relation holds even for a generalised LMF model with heterogenous strategies~\cite{SatoJSP2023}.

		In the pioneering work~\cite{LMF}, they provided a scatterplot between $\alpha$ and $\gamma$ by analysing an off-book market dataset as a proxy for hidden orders. We extracted the data in the figure in Ref.~\cite{LMF} and plot it as Fig.~\ref{fig:LMFPaper} with the red regression line added. This data shows two points: 
		\begin{enumerate}
			\item The theoretical line passes roughly through the centre of the data points, suggesting the minimum qualitative consistency between the data and theory. 
			\item At the same time, the theoretical line does not exhibit a good fit in explaining the {``variations in the measured values"}. Indeed, the red regression line has the coefficient of the slope $0.07$, which is far from the theoretical coefficient\footnote{
			They state, {``As a stronger test, one might hope that variations in measured values of $\alpha$ might predict variations in measured values of $\gamma$. The model fails this test"} in Ref.~\cite{LMF}.
			}. Perhaps, this might be partly due to their ``improper proxy''\footnote{
				They state, {``Because we lack the proper data to test the model, we have used an imperfect proxy to test the model''} in Ref.~\cite{LMF}.
			} and the smallness of the sample size.
		\end{enumerate} 
		It should be noted that Refs.~\cite{Vaglica,Bershova} showed that $\alpha\approx 1.5$ is empirically obtained on the basis of the aggregated distribution, suggesting the LMF prediction is consistent at least qualitatively.
		However, to establish the LMF quantitative prediction~\eqref{eq:review_LMP_quantPred}, it is necessary to solve the second problem by analysing a large and proper dataset.

	\subsection{Qualitative prediction and the corresponding empirical evidence}
		While the quantitative prediction~\eqref{eq:review_LMP_quantPred} is interesting, it can be another option to examine a rather weaker prediction by the LMF model. According to Ref.~\cite{Toth2015}, let us decompose the ACF $C(\tau)$:
		\begin{equation}
			C(\tau) := C_{\rm same}(\tau) + C_{\rm other}(\tau).
		\end{equation}
		Here $C_{\rm same}(\tau)$ is the contribution where the same trader issues orders at $t$ and $t+\tau$, whereas $C_{\rm other}(\tau)$ is the contribution where two distinct traders issue orders at $t$ and $t+\tau$. If the order-splitting hypothesis is correct, the following relationship is expected to hold: 
		\begin{equation}
			\left|C_{\rm same}(\tau)\right| \gg \left|C_{\rm other}(\tau)\right|, \>\>\> 
			\tau \gg 1.
			\label{eq:review_LMP_qualitativePred}
		\end{equation}
		We call this relationship~\eqref{eq:review_LMP_qualitativePred} ``the qualitative prediction of the LMF model", in comparison to the qualitative prediction~\eqref{eq:review_LMP_quantPred}, in this report. Ref.~\cite{Toth2015} addressed this problem and showed that the quantitative prediction~\eqref{eq:review_LMP_qualitativePred} actually holds in their dataset. This is the best empirical evidence supporting the order-splitting hypothesis, to the best of our knowledge.

	\subsection{Goal of this report}
		The excellent evidence in \cite{Toth2015} suggests the strong relevance of the order-splitting hypothesis as the microscopic origin of the LRC, at least on the qualitative level~\eqref{eq:review_LMP_qualitativePred}. At the same time, it is remarkable that the LMF model further provides the quantitative prediction~\eqref{eq:review_LMP_quantPred}, which is much stronger than the qualitative prediction~\eqref{eq:review_LMP_qualitativePred}. This relationship~\eqref{eq:review_LMP_quantPred} is obviously appealing. However, there has been no systematic and solid evidence to support this prediction at the quantitative level.
		
		The goal of this report is to examine and establish the quantitative prediction~\eqref{eq:review_LMP_quantPred} by analysing our large dataset on the TSE market. To prove the relationship~\eqref{eq:review_LMP_quantPred}, it is sufficient to draw a scatterplot between $\alpha$ and $\gamma$ with a sufficiently-large sample size. Therefore, we basically proceed with our data analysis in the following three steps (see also Appendix.~\ref{sec:app:summary_technicalproblems} for the summary of the technical problems to be solved): (1) measurement of the microscopic parameter $\alpha$, (2) measurement of the macroscopic parameter $\gamma$, and (3) drawing the scattter plot between $\alpha$ and $\gamma$.

\section{Measurement of the the metaorder-length distribution}\label{sec:Measure_MetaOrder_alpha}
	Here we classify the random and order-splitting strategies in terms of the market orders to finally measure the microscopic parameter $\alpha$ in the metaorder distribution $P(L)\propto L^{-\alpha-1}$ for large $L$. 
	
	\subsection{Measurement of metaorder lengths for individual traders}\label{sec:def_metaorderLength}
		We first define the metaorder series at the level of individual traders. Basically, we follow the rule described in Sec.~\ref{sec:math_notation} to extract the order-sign sequence $\{\eps_{k}^{(i)}\}_k$ for the $i$-th trader and to construct the corresponding run sequences $\{L_{k}^{(i)}\}_k$. The run sequences are regarded the {\it metaorder-length sequences} in this report.  
		
		For a practical reason, however, we introduce one exceptional rule: if the time interval between two successive orders are sufficiently longer, we regard that two orders belong to different metaorders according to Ref.~\cite{Donier}. This rule is introduced to avoid overestimation of unrelated orders. For example, let us consider the case where a trader submits ten buy orders within a day, stops orders for one month, and then submits ten buy orders. It is not realistic to assume that the metaorder length is twenty because the one-month resting seems too long. We expect this exceptional rule will reduce the risk of such overestimation. In this report, we set this time threshold to be one business day. 

	\subsection{Strategy clustering: random vs. order-splitting traders}
		We next identify the {\it order-splitting traders} (STs) at the level of individual traders. Our basic idea is to apply the binomial test in statistics to define the {\it random traders} (RTs) and then define STs as non-RTs. The details of our strategy-clustering methods and the corresponding results are described below. 
		
		\subsubsection{Methods: the binomial test}\label{sec:binomial_test}
			Let us define the RTs by the binomial test as follows: If a trader $i$ randomly issues market orders, it is expected that the sign sequence $\{L^{(i)}_k\}_k$ is generated according to the symmetric Bernoulli process. In other words, the sign sequence obeys the rule
			\begin{equation}
				P\left(\eps^{(i)}_{k} = +1 \>|\>  \eps^{(i)}_{k-1}, ..., \eps^{(i)}_1\right) = \frac{1}{2}
				\label{eq:random}
			\end{equation}
			for any $k\geq 1$. 

			On the basis of this picture, we set the following null hypothesis:  
			\begin{equation}
				H_0: \mbox{ the sign sequence of the trader $i$ obeys the symmetric Bernoulli process.}
			\end{equation}
			This hypothesis is examined by the one-sided binomial test with the significance level $\theta:=0.01$ as follows: 
			Let us consider the reduced order-sign sequence $\{\eps^{(i)}_k\}_{k}$ of the $i$th trader. The total number of their market orders is given by $N^{(i)}_{\rm MO}:=|\{\eps^{(i)}_k\}_{k}|$ and the corresponding run-length sequence is given by $\{L^{(i)}_k\}_k$. We here focus on the total number of runs defined by $N_{\rm run}^{(i)}:=|\{L^{(i)}_k\}_k|$. If the null hypothsis $H_0$ is correct, the total number of runs $N_{\rm run}^{(i)}$ must obey the binomial distribution, 
			\begin{equation}
				P(N_{\rm run}^{(i)}) =  \frac{1}{2^{N^{(i)}_{\rm MO}-1}}\binom{N^{(i)}_{\rm MO}-1}{N_{\rm run}^{(i)}}.
			\end{equation}
			We thus apply the one-sided binomial test to testify the null hypothesis $H_0$. 
			If this null hypothesis is rejected, we classify the trader $i$ as a ST (or just an ST for short), such that $i\in \Omega_{\ST}$ with the set of the STs $\Omega_{\ST}$; otherwise, the trader $i$ is classified as an RT (i.e., $i\in \Omega_{\RT}$ with the set of the RTs $\Omega_{\RT}$). The first-kind error (the false-positive rate) is controlled in our statistical test, and the clustering for the STs 
			
		\subsubsection{Results 1: the existence of the order-splitting traders}
		\label{sec:ResultClassification}
			\begin{figure}
				\centering
				\includegraphics[width=170mm]{"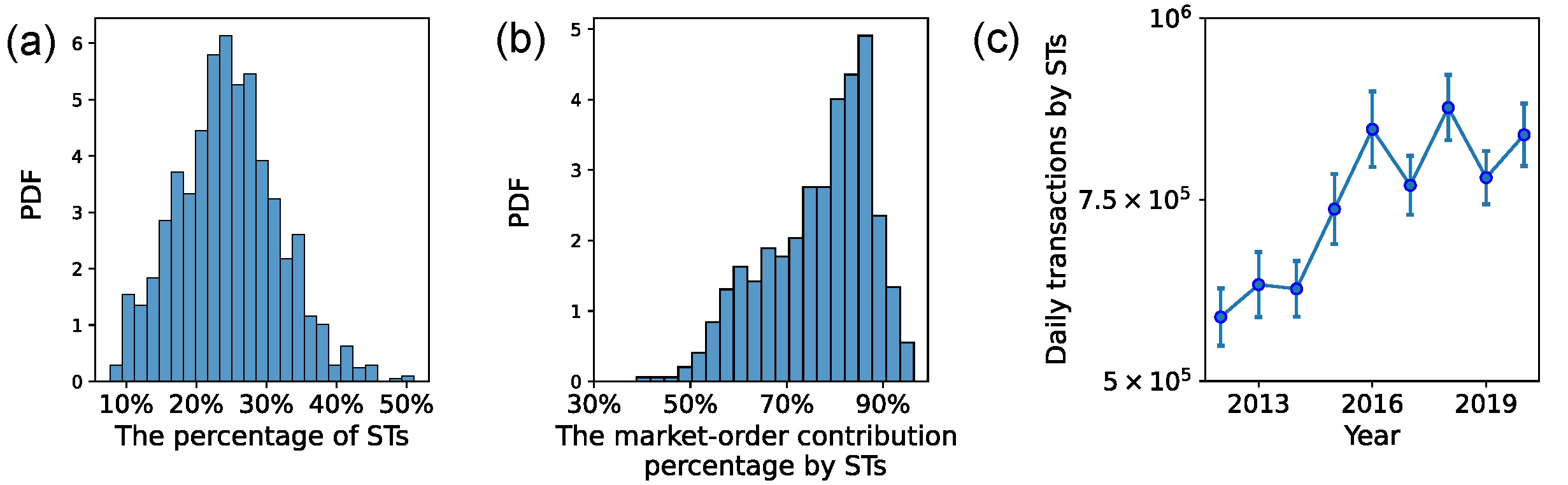"}
				\caption{
						Summary statistics of STs. 
						(a)~Empirical distribution of the percentage of STs in each market. Approximately 25\% ~(10-50\%) of traders are classified as STs in each market. 
						(b)~Empirical distribution of the market-order contribution percentage by STs. The 80\% of the total market orders are typically submitted by STs. These two figures suggest that STs dominantly contribute to the market orders, while their number is relatively fewer than that of RTs. 
						(c)~Yearly plot of average daily transaction numbers by STs, showing the growing presence of STs.
				}
				\label{fig:ratio_OSTs}
			\end{figure}
			
			Let us show the overview of our clustering results. We applied the clustering algorithm in Sec.~\ref{sec:binomial_test} to all traders for all stock every year (i.e., one datapoint $\Gamma$) to obtain $\Omega_{\ST}$ and $\Omega_{\RT}$. We can define the ratio of the STs $|\Omega_{\ST}|/|\Omega_{\TR}|$ for each datapoint. We first show the empirical distribution of the STs percentage as Fig.~\ref{fig:ratio_OSTs}(a). The typical percentage of the STs are given by 25\%, showing the direct evidence of the presence of the STs in our dataset. 

			Interestingly, while the number of the STs are typically less than that of the RTs, the STs typically exhibits the dominant contribution to the market orders. To show this character, let us define the market-order contribution percentage by the STs as the ratio of the number of the market orders issued by the STs to the total number of the market orders. Figure~\ref{fig:ratio_OSTs}(b) shows the empirical distribution of the market-order contribution percentage by the STs, illustrating that the STs typically contribute 80\% to the total market orders. In addition, the presence of STs shows a tendency to increase over the years (see Fig.~\ref{fig:ratio_OSTs}(c)).

		\subsubsection{Results 2: metaorder-length distributions}
			\label{sec:metaorder_alpha_estimation}
			\begin{figure}
				\centering
				\includegraphics[width=100mm]{"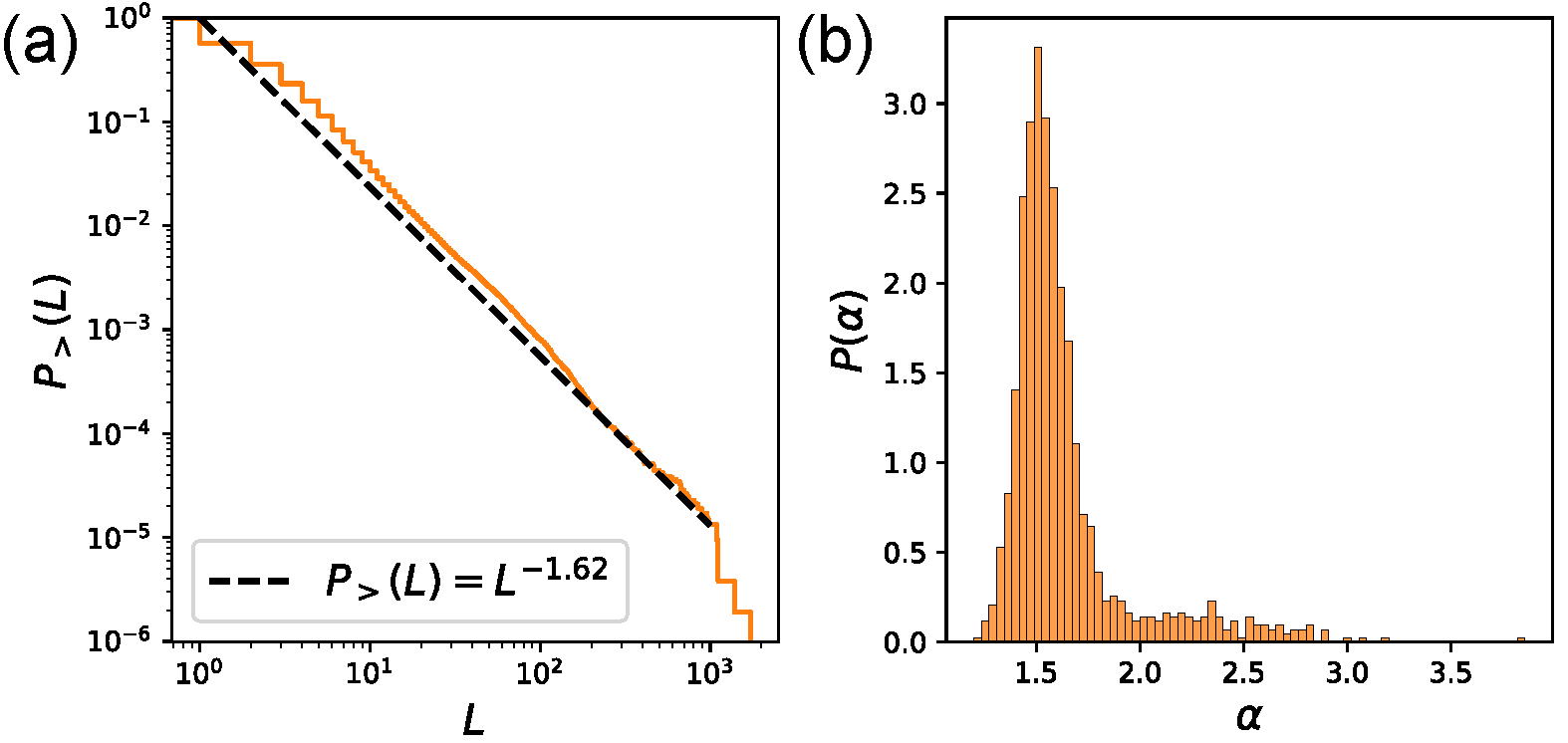"}
				\caption{
					 Characters of metaorder-length distribution on STs in each markets.
						(c)~The aggregated metaorder-length CCDF for all STs for Toyota Motor Corporation in 2020 as a typical example. The metaorder-length CCDF obeys the power law such that $P_>(L)\sim L^{-\alpha}$.
						(d)~The empirical PDF of the power-law exponents $\alpha$ in our whole dataset. The exponent $\alpha$ was measured by Clauset's algorithm~\cite{Clauset2009,Alstott2014} across all the markets. Typically, $\alpha$ distributes within $1<\alpha<2$, consistently with the standard assumption for the LMF model. 
				}
				\label{fig:agg_meta}
			\end{figure}

			We then study the metaorder-length CCDF for the STs (see also Appendix~\ref{app:metaorder_RTs} for the clustering results of RTs as a reference). Let us consider the joint run-length sequences for STs and the corresponding metaorder-length CCDF:
			\begin{equation}
				\left\{L^{\ST}_k\right\}_{k} := 
				\bigcup_{i \in \Omega_{\ST}} \left\{L_k^{(i)}\right\}_{k}, \>\>\> 
				P_>\left(L^{\ST}\right) := \frac{N_>(L^{\ST})}{\left|\{L^{\ST}_k\}_{k}\right|}, \>\>\> 
				N_>(L^{\ST}) := \int_{L^{\ST}}^\infty dy\sum_{k} \delta\left(y-L^{\ST}_k\right).
			\end{equation}
			The empirical metaorder distribution for STs is plotted in Fig.~\ref{fig:agg_meta}(a) for Toyota 2020, showing the power law $P_>(L)\approx L^{-\alpha}$. We confirm that this character is robustly observed even for other datapoints. 

			The empirical PDF $P(\alpha)$ of the power-law exponent $\alpha$ is shown in Fig.~\ref{fig:agg_meta}(b). Approximately 90\% of the stocks have power-law exponent $\alpha < 2$ in our dataset. This finding is consistent with the standard assumption that $\alpha < 2$ in the LMF model. The power-law exponent $\alpha$ is estimated by Clauset's algorithm~\cite{Clauset2009,Alstott2014} as one of the established statistical estimation methods. According to Ref.~\cite{Clauset2009}, the estimation errors in the power-law exponent are generally small, at least compared with the errors in another power-law exponent $\gamma$. We thus ignore the estimation errors of the power-law exponent $\alpha$ throughout this report.

		\subsubsection{Parameter estimation for the LMF model}
			Here we describe our method to estimate the parameters for the LMF model. 

			\begin{itemize}
				\item $N_{\ST}$: the total number of the active STs, trading at least one thousand times in the year, is estimated as the yearly average number of all the STs: 
				\begin{equation}
					N_{\ST} = \frac{1}{D_{\rm year}} \sum_{i \in \Omega_{\ST}} D^{(i)},
				\end{equation} 
				where $D_{\rm year}$ is the total number of the business days in that year, and $D^{(i)}$ is the total number of active days by the $i$-th ST. We used $N_{\ST}$ as a proxy for the parameter calibration of $N_{\TR}$ because the influence of inactive traders is expected to be negligible on the empirical autocorrelation function. In addition, it is numerically known that the asymptotic behaviour of the autocorrelation function is robust regarding the total number of traders $N_{\TR}$~\cite{BouchaudText}, and, thus, the technical details of the parameter calibration of $N_{\TR}$ are expected to be insensitive to the final results.

				\item $N_{\epsilon}$: we substitute the total number of market orders for the stock during the year into $N_{\epsilon}$.
								
				\item $\alpha$: the power-law exponent of the metaorder length PDF $P_{\ST}(L)\propto L^{-\alpha-1}$ for large $L$. This power-law exponent is estimated by the Clauset algorithm as described in Sec.~\ref{sec:metaorder_alpha_estimation}.

			\end{itemize}
			This parameter estimation method was used for the numerical simulations in Sec.~\ref{sec:measure_gamma}.

\section{Measurement of the sign autocorrelation function}\label{sec:measure_gamma}
	Here we describe the measurement of the power-law exponent $\gamma$ in the sign ACF $C(\tau)\propto \tau^{-\gamma}$ for large $\tau$. Our method is composed of three steps: (1) application of the naive estimator $\gamma_{\naive}$ by the nonlinear least squares (NLLS) to the empirical ACF or power-spectral density (PSD), (2) the construction of an unbiased estimator based on the LMF model, and (3) application of the unbiased estimator to obtain the final $\gamma_{\unbiased}$. Let us explain these steps one by one. 

	\subsection{Estimations by the nonlinear least squares}
		We employed two NLLS estimation methods based on the ACF and PSD for our statistical analyses. Both methods show similar and consistent results, implying the robustness of our analyses. While there are many sophisticated estimation methods (such as the estimation based on the detrended fluctuation analysis (DFA)~\cite{Lillo2004}), we employ this simple method because we find that the NLLS estimation has the consistency for infinite sample size and has less bias for finite sample size than other methods.

		\subsubsection{Estimation based on the sample ACF}
			Let us first describe the measurement method based on the sample ACF. The basic idea is to apply the power-law fitting to the sample ACF $C_{\sample}(\tau)$, such that 
			\begin{equation}
				C_{\sample}(\tau):=\frac{1}{N_{\eps}-\tau}\sum_{t=1}^{N_{\eps}-\tau} \eps(t)\eps(t+\tau) \propto \tau^{-\gamma_{\naive}^{\acf}},
			\end{equation}
			where the superscript $\acf$ signifies the ACF estimator. 
			The detailed implementation is described in Appendix.~\ref{sec:app:NLLS_ACF}.
			The theoretical advantage of this method is that the sample ACF is expected to converges to the true ACF for the infinite sample size: 
			\begin{equation}
				\lim_{N_{\eps} \to \infty}C_{\sample}(\tau) = \left<\eps(t)\eps(t+\tau)\right> = C(\tau)
				\label{eq:ergodicity_ACF}
			\end{equation}
			under the ergodicity assumption. In general, ergodicity is a weak assumption irrelevant to the underlying microscopic dynamics (in our case, the microscopic dynamics are assumed to be governed by the LMF model). Therefore, it is expected that the NLLS estimator has the consistency for general setups as shown in Sec.~\ref{sec:NLLS_consistency}. 
		
		\subsubsection{Estimation based on the sample PSD}
			Another method we employed is based on the PSD. The basic idea is to utilise the one-to-one correspondence between the PSD $S(\omega)$ and the ACF $C(\tau)$, guaranteed by the Wiener-Khinchin theorem,
			\begin{equation}
				S(\omega) = \int_{-\infty}^\infty C(\tau)e^{2\pi i\omega \tau}d\tau.
				\label{eq:WK-theorem}
			\end{equation}
			Considering the integral identity~\cite{Tauber} for $\gamma \in (0,1)$
			\begin{equation}
				\int_{-\infty}^\infty e^{2\pi i \omega \tau}|\tau|^{-\gamma}d\tau = 2^\gamma \pi^{\gamma-1}\Gamma(1-\gamma)\sin \frac{\pi\gamma}{2}|\omega|^{\gamma-1},
			\end{equation}
			if the sample PSD obeys the power-law asymptotics for small $\omega$ 
			\begin{equation}
				S_{\sample}(\omega) \propto \omega^{-H},
			\end{equation}
			the hurst exponent $H$ is related to $\gamma$ as $H=1-\gamma$. In other words, the NLLS estimator $\gamma_{\naive}^{\psd}$ is defined by
			\begin{equation}
				\gamma_{\naive}^{\psd} = 1- H,
			\end{equation}
			where $H$ is determined by the NLLS method.	The superscript $\psd$ signified the PSD estimator. As with the sample ACF, the sample PSD converges to the true PSD under the ergodicity assumption. Therefore, this estimation is expected to be robust regarding the consistency (see Sec.~\ref{sec:NLLS_consistency}). The detailed implementation is described in Appendix~\ref{sec:app:NLLS_PSD}.
			
			\subsubsection{Comparison between the ACF and PSD methods}
				We discuss theoretical differences between the ACF and PSD methods for comparison. The NLLS fitting sometimes provides negative values of $\gamma < 0\Longleftrightarrow H > 1$ due to methodological artefacts. Negative $\gamma$ implies monotonic increasing of the ACF for large $\tau$, which does not make sense. We excluded such datapoints because of the obvious failure of the estimation\footnote{The total number of the datapoints was 16 which needs exceptional handling with $\gamma<0$.}.

				The PSD estimator is theoretically valid only for $H \in (0,1)$, or equivalently, $\alpha \in (1,2)$. This fact means the estimation fails for $\alpha>2$ in principle, even for infinite observations. Also, the estimation accuracy tends to be worse near the critical point $\alpha=2$. 
				
				These disadvantages contrast with the ACF method, which is expected to work for any $\alpha$ in principle for infinite observations and provide only positive $\gamma$. However, the PSD method is broadly used to estimate the Hurst exponent and is a realistic option for the statistical estimation of $\gamma$. Indeed, as for the LMF simulations, we find that the overall bias due to the finite sample size was less in the PSD method than in the ACF method (see the value of $\beta_1$ in Sec.~\ref{sec:bias_finite_N}).

	\subsection{Consistency and biasedness of the NLLS estimator}\label{sec:NLLS_consistency}
	\begin{figure}
		\centering
		\includegraphics[width=170mm]{"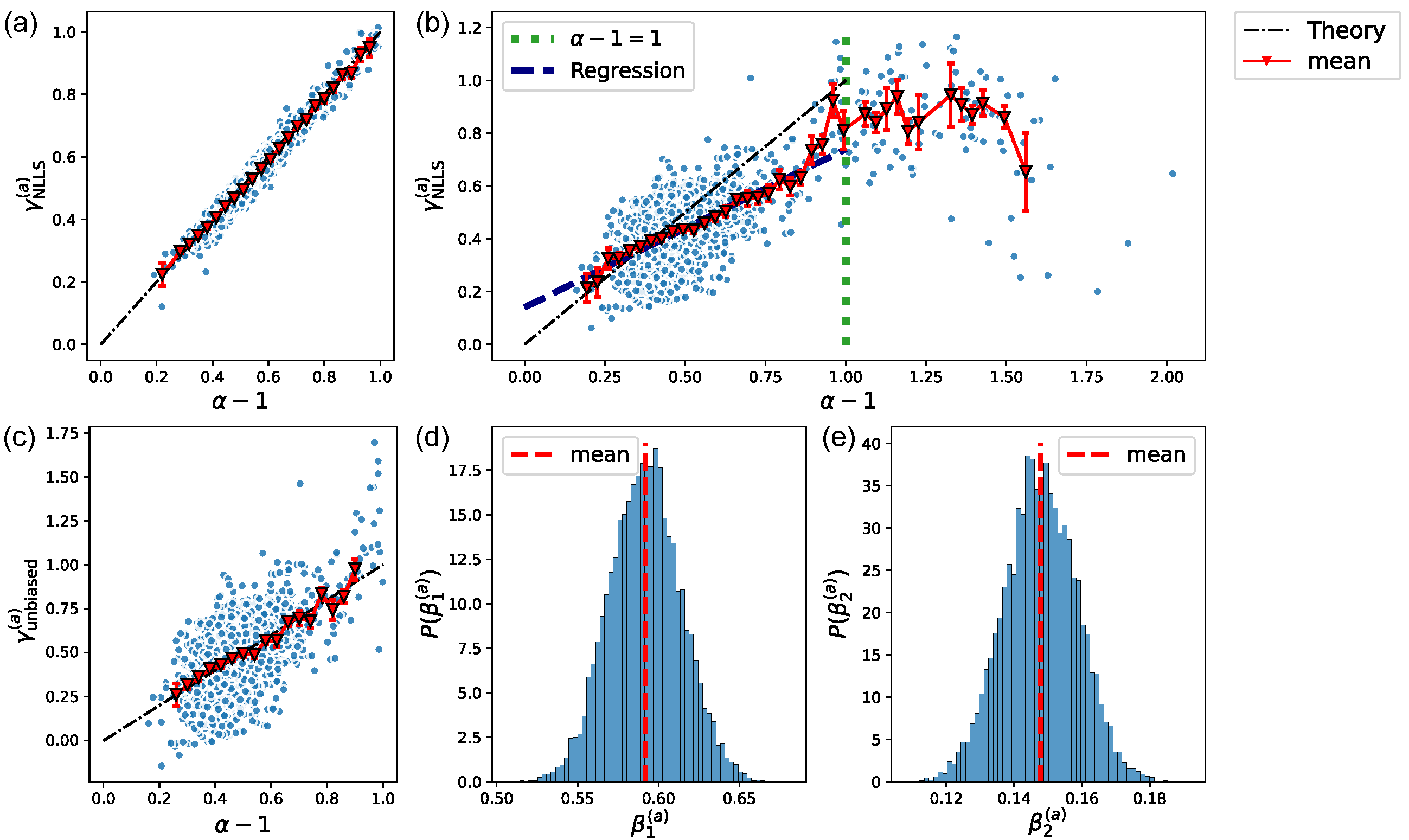"}
		\caption{
			Numerical simulations of the LMF model to test consistency and biasedness of the NLLS estimator $\gamma_{\naive}^{\acf}$ based on the ACF method. 
			(a)~Consistency of the ACF-NLLS estimator $\gamma_{\naive}^{\acf}$. The theoretical formula $\gamma_{\naive}^{\acf}=\alpha-1$ holds for a sufficiently-large sample size $N_\eps = 10^8$, supporting the consistency of the NLLS estimator $\gamma_{\naive}^{\acf}$. The parameter sets $(N_{\ST},\alpha)$ are based on the measurement of our dataset.  
			(b)~For realistic sample sizes $N_\eps \lesssim 10^7$, the ACF-NLLS estimator $\gamma_{\naive}^{\acf}$ exhibits biasedness due to the finite sample size. We have simulated the LMF model by assuming that the parameter sets $(N_{\ST},N_{\eps},\alpha)$ are identical to those measured in our dataset. We find that the theoretical formula $\gamma_{\naive}^{\acf}=\alpha-1$ does not hold due to the finite sample size. The deviation from the theoretical line is particularly serious for large $\alpha-1>1$, showing the non-uniform convergence of the NLLS estimator. Based on this empirical finding, we focus on the range $\alpha - 1 \in (0,1)$ and apply the linear regression~\eqref{eq:linear_regression} to obtain the navy line. 
			(c)~Check of the numerically-constructed unbiased ACF estimator $\gamma_{\unbiased}^{\acf}$. The unbiased estimator is constructed by $\gamma_{\unbiased}^{\acf}:=(\gamma_{\naive}^{\acf}-\beta_2^{\acf})/\beta_1^{\acf}$. As expected, the theoretical formula $\gamma_{\unbiased}^{\acf}\approx \alpha-1$ holds even for the realistic sample sizes.
			(d and e)~Parameter distribution of $\beta_1^{\acf}$ and $\beta_2^{\acf}$, the coefficients of the regression formula~\eqref{eq:linear_regression} in constructing the unbiased estimator $\gamma_{\unbiased}^{\acf}$. The means of $\beta_1^\acf$ and $\beta_2^\acf$ were given by $0.592$ and $0.147$, respectively.
		}
		\label{fig:simresult-ACF}
	\end{figure}
	\begin{figure}
		\centering
		\includegraphics[width=170mm]{"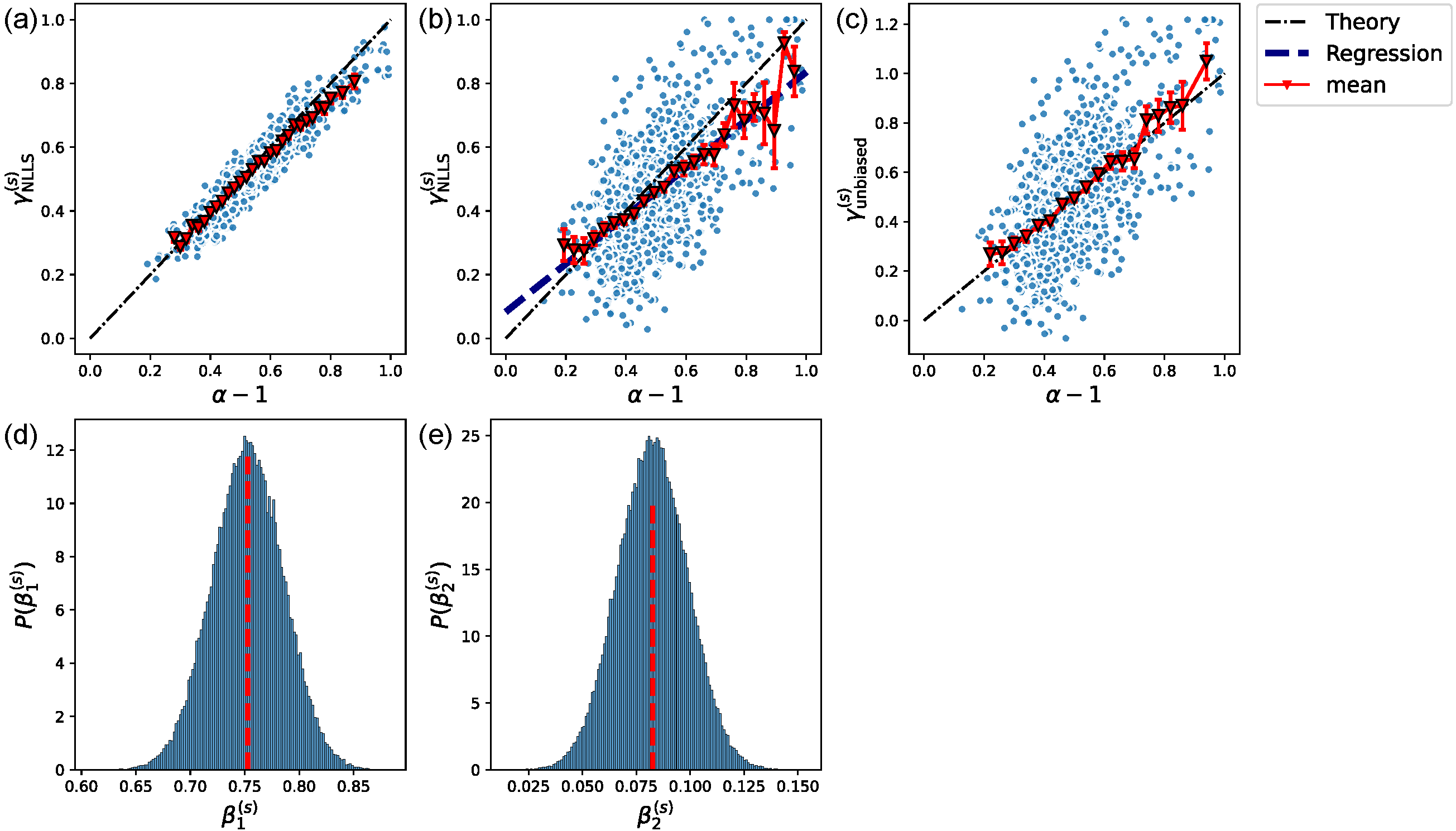"}
		\caption{
			Numerical tests for consistency and biasedness of the NLLS estimator $\gamma_{\naive}^{\psd}$ based on the PSD method. The test was based on the LMF model.
			(a)~Consistency of $\gamma_{\naive}^{\psd}$ for a sufficiently-large sample size $N_\eps = 10^8$. The parameter sets $(N_{\ST},\alpha)$ are based on the measurement of our dataset.  
			(b)~For realistic sample sizes $N_\eps \lesssim 10^7$, the PSD-NLLS estimator $\gamma_{\naive}^{\psd}$ exhibits biasedness due to the finite sample size. The LMF model was simulated by assuming that the parameter sets $(N_{\ST},N_{\eps},\alpha)$ are identical to those measured in our dataset. We applied the linear regression~\eqref{eq:linear_regression} for the range $\alpha - 1 \in (0,1)$ to obtain the navy line. 
			(c)~Confirmation of the numerically-constructed unbiased PSD estimator $\gamma_{\unbiased}^{\psd}$. The unbiased estimator is constructed by $\gamma_{\unbiased}^{\psd}:=(\gamma_{\naive}^{\psd}-\beta_2^{\psd})/\beta_1^{\psd}$. As expected, $\gamma_{\unbiased}^{\psd}\approx \alpha-1$ holds even for the realistic sample sizes.
			(d and e)~Parameter distribution of $\beta_1^{\psd}$ and $\beta_2^{\psd}$, the coefficients of the regression formula~\eqref{eq:linear_regression} in constructing the unbiased estimator $\gamma_{\unbiased}^{\psd}$. The means of $\beta_1^\psd$ and $\beta_2^\psd$ were given by $0.753$ and $0.083$, respectively.
		}
		\label{fig:simresult-PSD}
	\end{figure}

		In statistics, the {\it consistency} and {\it unbiasedness} are two of the desirable characters of any statistical estimator. Here we numerically confirm these characters of the NLLS estimator based on the LMF model (see Figs.~\ref{fig:simresult-ACF} and \ref{fig:simresult-PSD} for the ACF and PSD methods, respectively). While the NLLS estimator has the consistency at least numerically, unfortunately, it does not have the unbiasedness. This problem is heuristically solved in Sec.~\ref{sec:construct_unbiasedEstimator} by appropriate construction of an unbiased estimator.

		\subsubsection{Consistency for the infinite sample size}
			Any estimator $T_{n}$ is called consistent if the estimated value converges to the true value $\theta$ for the infinite sample size $n \to \infty$: $\lim_{n\to \infty} T_n = \theta$. We have numerically confirmed the consistency of the NLLS estimator: 
			\begin{equation}
				\lim_{N_{\eps}\to \infty}\gamma_{\naive} = \gamma.
				\label{eq:NLLS_consistency}
			\end{equation}
			To confirm this consistency~\eqref{eq:NLLS_consistency}, we have numerically generated the order-sign sequences by the LMF model with realistic parameters of our dataset: we have measured the model parameter set $(N_{\ST}, \alpha)$ for all sample points according to Sec.~\ref{sec:metaorder_alpha_estimation} except for $N_\eps$. For $N_{\eps}$, we employed $N_{\eps}=10^{8}$ because realistic values $N_{\eps}\lesssim  10^7$ are not sufficient to confirm the consistency~\eqref{eq:NLLS_consistency}.  
			Figures~\ref{fig:simresult-ACF}(a) and \ref{fig:simresult-PSD}(a) illustrate our numerical simulation, showing that the NLLS estimator $\gamma_{\naive}$ numerically agrees with the theoretical formula $\gamma=\alpha-1$. This numerical evidence supports the consistency of the NLLS estimator. Note that the consistency of the NLLS estimator is theoretically reasonable because the sample ACF (PSD) converges to the true ACF (PSD) for the infinite sample size under the assumption of ergodicity. Note that the PSD method works slightly worse near $\alpha\approx 2$ than the ACF method (see Fig.~\ref{fig:simresult-PSD}(a)) because $\alpha=2$ is the critical point beyond which the PSD method fails to estimate $\alpha$ in principle.

		\subsubsection{Bias for the finite sample size}\label{sec:bias_finite_N}
			Any estimator $T_n$ is called unbiased if the expectation of the estimator is equivalent to the true value $\theta$ for the finite sample size $n < \infty$: $\la T_n\ra=\theta$. Unfortunately, we have numerically confirmed that the NLLS estimator does not have the unbiasedness:
			\begin{equation}
				\la \gamma_{\naive}\ra \neq \gamma \>\>\> \mbox{ for finite }N_{\eps}.
			\end{equation}
			To confirm this character, we have numerically generated the order-sign sequences by the LMF model with realistic parameters of our dataset: we measured the model parameter set $(N_{\ST}, \alpha, N_{\eps})$ for all sample points according to the method in Sec.~\ref{sec:metaorder_alpha_estimation}. Under the measured parameter sets, we numerically performed the  Monte Carlo simulations of the LMF model. The scatterplots are generated 100 times as IID realisations, and take their ensemble average based on the the bootstrap method to draw the final scatterplot.

			Under realistic parameter sets, as shown in Fig.~\ref{fig:simresult-ACF}(b) and \ref{fig:simresult-PSD}(b), we find the systematic deviation between the theoretical line $\gamma=\alpha-1$ and the numerical datapoints. This suggests the NLLS estimator has the finite-sample-size bias. In addition, we find that the convergence speed is very slow for finite $N_{\eps}$ and is not even uniform in terms of $\alpha$. It is reasonable that the convergence is non-uniform in terms of $\alpha$. Indeed, the decay speed of the ACF is so fast for larger $\alpha$ that the power-law part of the ACF cannot be observed for a wide range of $\tau$. For this practical reason, we have restricted our analyses to the range $\alpha \in (1,2)$, which agrees with the standard assumption of the LMF model.

			By focusing on the range $\alpha \in (1,2)$, let us apply the linear regression between $\gamma_{\naive}$ and $\alpha$ as shown in Fig.~\ref{fig:simresult-ACF}(b) according to the formula: 
			\begin{align}
				\gamma_{\naive} = \beta_1 (\alpha-1)+\beta_2, \label{eq:linear_regression}
			\end{align}
			where $\beta_1$ and $\beta_2$ are regression coefficients. This relation is used to numerically construct an unbiased estimator (see Figs.~\ref{fig:simresult-ACF}(c) and ~\ref{fig:simresult-PSD}(c) for the ACF and PSD methods, respectively) as shown in Sec.~\ref{sec:construct_unbiasedEstimator}.

			If the NLLS were an unbiased estimator, the relations $\la\beta_1\ra\approx 1$ and $\la \beta_2\ra\approx 0$ would hold. To test these relations, we repeated the numerical simulations of the LMF model and linear regressions~\eqref{eq:linear_regression} to obtain the empirical histograms of $\beta_1$ and $\beta_2$ (see Figs.~\ref{fig:simresult-ACF} (d) and (e) for the ACF method and Figs.~\ref{fig:simresult-PSD} (d) and (e) for the PSD method). We numerically find that $\la \beta_1^\acf \ra = 0.592$ and $\la \beta_2^\acf\ra = 0.147$ for the ACF method and $\la \beta_1^\spectra \ra = 0.753$ and $\la \beta_2^\spectra\ra = 0.083$ for the PSD method in our simulations. These values clearly show the biasedness of the NLLS estimator due to the finite sample size. To solve this finite-sample-size bias problem, we will approximately construct a numerical unbiased estimator in Sec.~\ref{sec:construct_unbiasedEstimator}. 

		\subsubsection{Other methods: the detrended fluctuation analysis}
			\begin{figure}
				\centering
				\includegraphics[width=70mm]{"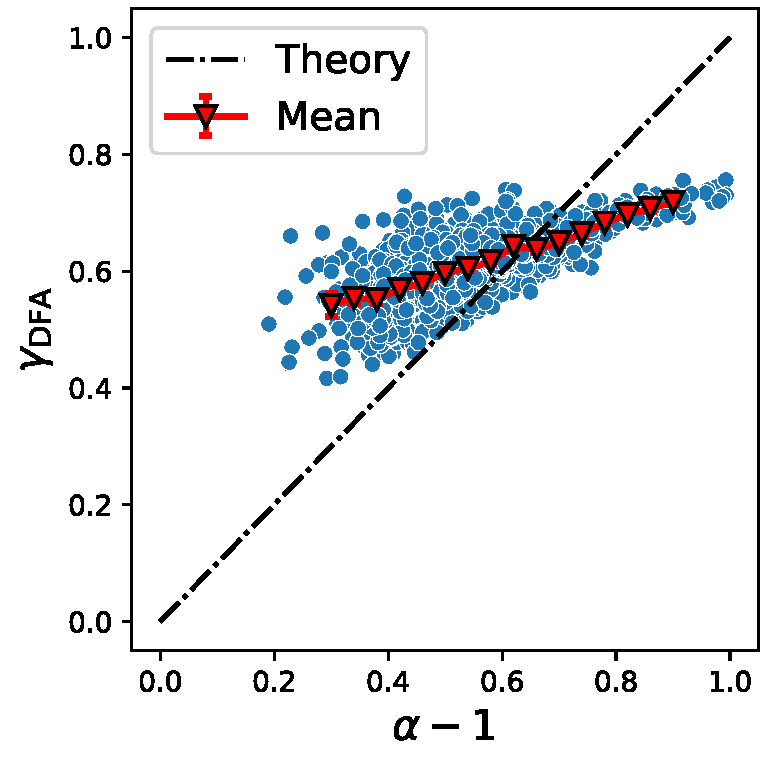"}
				\caption{
					Inconsistency of the DFA estimator $\gamma_{\DFA}$ even for the LMF model. We used the same parameters $(N_{\ST},\alpha)$ as those measured in our dataset and set $N_\eps = 10^8$. The theoretical relationship $\gamma_{\DFA}=\alpha-1$ does not hold even for large sample size, rejecting the numerical consistency of the DFA estimator $\gamma_{\DFA}$ at least under a realistic computational resource.
			}\label{fig:DFA}
			\end{figure}

			There are several other methods to estimate the power-law exponent $\gamma$, and one of the famous methods is based on the Hurst exponent with the detrended fluctuation analysis (DFA)~\cite{Hurst}. Indeed, some researchers claim that the DFA analysis provides much better results than the NLLS estimation \cite{Lillo2004,LMF}. In this report, the estimated exponent by the DFA is called the DFA estimator and is denoted by $\gamma_{\DFA}$.

			We do not use the DFA estimator because we numerically find a serious problem of the DFA estimator in terms of the finite-sample-size bias. We numerically generated the order-sign sequences by the LMF model and measured $\gamma_{\DFA}$ to obtain Fig.~\ref{fig:DFA}. The sample size is set to be $N_{\eps}=10^8$ because the NLLS estimator showed the consistency under this sample size.

			The DFA estimator is numerically implemented by using the referred Python package provided by Ref.~\cite{DFA}. Remarkably, the DFA estimator $\gamma_{\DFA}$ systematically deviates from the theoretical line~\eqref{eq:LMF_prediction}. Since the theoretical line~\eqref{eq:LMF_prediction} is the exact solution for the LMF model, this deviation signifies the serious bias of the DFA estimator. Unfortunately, within our computational resource, we could not even confirm the consistency of the DFA estimator for larger sample size. 

			We are not sure about its crucial reason currently, but one of the potential reasons might be related to the stronger statistical assumption required by the DFA estimator. While the consistency of the DFA estimator was recently proved for the fractional Brownian motion~\cite{Consistency_DFA}, it is non-trivial whether the consistency of the DFA estimator is still kept even for systems not obeying the fractional Brownian motion. In our case, there is no solid reason why the sign sequence generated by the LMF model can be regarded as the fractional Brownian motion. On contrary, the NLLS estimator relies only on the ergodic assumption: the sample ACF converges to the true ACF for a large sample size. In this sense, the NLLS estimator requires weaker statistical assumptions than the DFA estimator. This might be a potential reason causing the difference between $\gamma_{\naive}$ and $\gamma_{\DFA}$.

			Here we do not claim inappropriateness of the DFA in general context. However, since our aim is to verify the quantitative prediction~\eqref{eq:LMF_prediction} based on the LMF model, we should use a less biased estimator in terms of the scatterplot between $\gamma$ and $\alpha$ at least for the LMF simulations. Thus, we do not use the DFA estimator in this report.

	\subsection{Numerical construction of an approximate unbiased estimator}\label{sec:construct_unbiasedEstimator}
		While the NLLS estimator numerically exhibits the consistency, it is biased for finite sample size with slow convergence speed. This problem should be solved before the direct verification of the LMF prediction~\eqref{eq:LMF_prediction}. In this subsection, we approximately construct an unbiased estimator based on the  LMF model. 

		Our idea for our unbiased estimator is based on our numerical observation of the scatterplot in Figs.~\ref{fig:simresult-ACF}(b) and \ref{fig:simresult-PSD}(b) for the ACF and PSD methods, respectively. For the numerical simulations of our LMF model, we numerically find that $\gamma_{\naive}$ follows the linear regression relation~\eqref{eq:linear_regression} for the range $\alpha \in (1,2)$ at least approximately. Since the relation~\eqref{eq:LMF_prediction} holds for the LMF model, the approximate relation $\gamma_{\naive}\approx \beta_1\gamma +\beta_2$ should holds between the NLLS estimator $\gamma_{\naive}$ and the true $\gamma$. Therefore, we numerically construct an unbiased estimator $\gamma_{\unbiased}$ as 
		\begin{equation}
			\gamma_{\unbiased} := \frac{\gamma_{\naive}-\beta_2}{\beta_1},\label{eq:modification}
		\end{equation}
		which exhibits the approximate unbiasedness, at least for the LMF simulations (see Figs.~\ref{fig:simresult-ACF}(c) and \ref{fig:simresult-PSD}(c) for the ACF and PSD methods, respectively), as 
		\begin{equation}
				\la\gamma_{\unbiased} \ra \approx \gamma.
		\end{equation}
		For the verification of the LMF prediction~\eqref{eq:LMF_prediction} in Sec.~\ref{sec:ScatterPlot}, we use this unbiased estimator $\gamma_{\unbiased}$.

\section{Verification of the LMF prediction}\label{sec:ScatterPlot}
	Let us proceed with the main result of this report: the direct verification of the LMF prediction~\eqref{eq:LMF_prediction}. We then discuss the relationship to previous works, possible future implications, and some open questions.

	\subsection{Scatterplot about the power-law exponent}
		\begin{figure}
			\centering
			\includegraphics[width=170mm]{"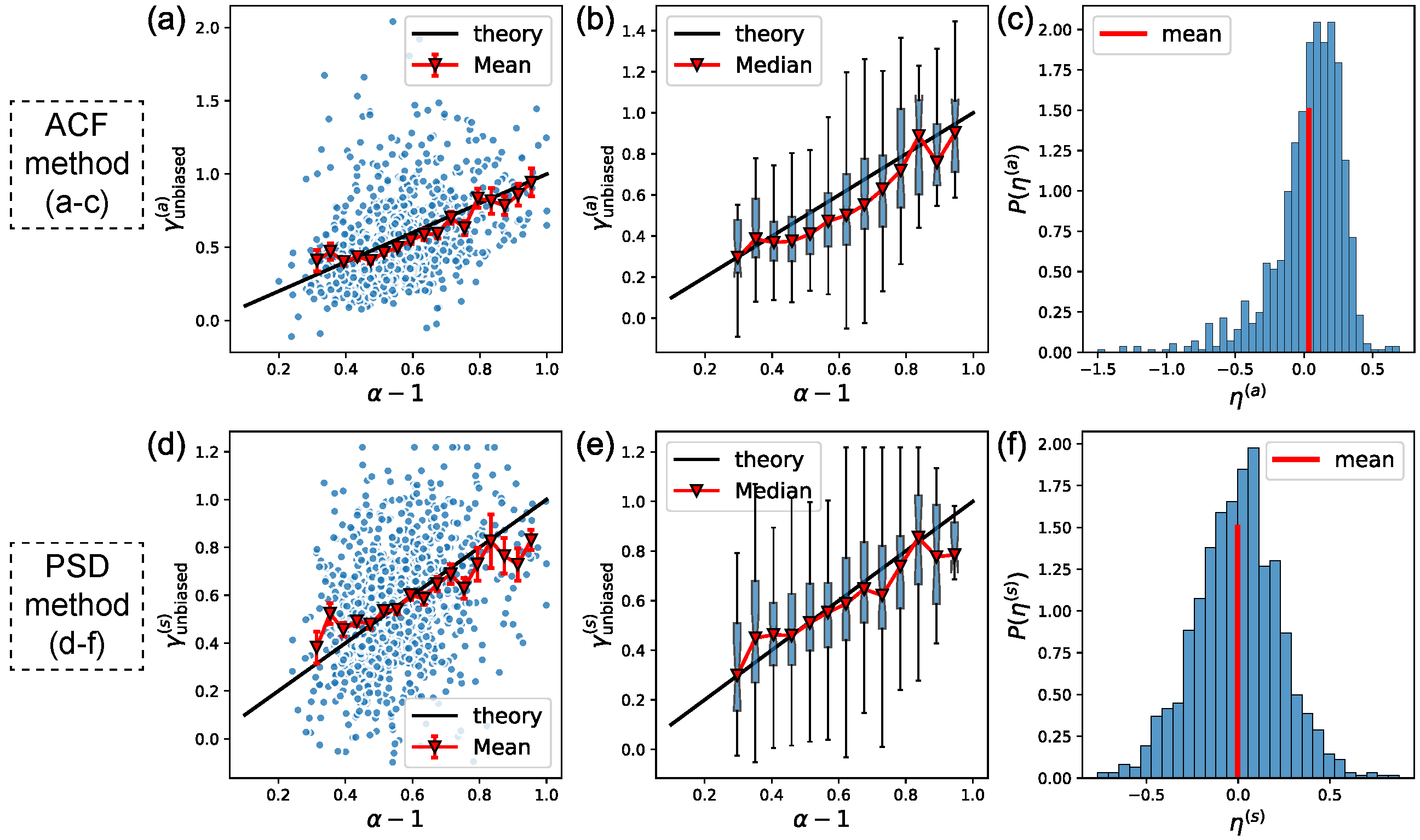"}
			\caption{
				Direct verification of the LMF prediction~\eqref{eq:LMF_prediction} based on the ACF method (a-c) and the PSD method (d-f). 
				(a, d)~Scatterplot between $\alpha$ and $\gamma_{\unbiased}$. We set bins along the $\alpha$ axis and the red line signifies the averages within the bins. 
				(b, e)~Box plot between $\alpha$ and $\gamma_{\unbiased}$, where statistical quantities (e.g., the first, second (median), and third quartiles) are calculated within the bins along the $\alpha$ axis.
				(c, f)~Empirical PDF of the errors $\eta:=\alpha-1-\gamma_{\unbiased}$. The average error is very small such that $\la \eta^{\acf}\ra= 0.03$ for the ACF method and $\la \eta^{\psd}\ra= 0.003$ for the PSD method, respectively.
			}
			\label{fig:scatterplot_empirical}
		\end{figure}

		For the verification of the LMF prediction~\eqref{eq:LMF_prediction}, we plot the scatterplot between $\alpha$ and $\gamma_{\unbiased}$ for $\alpha \in (1,2)$ (see Figs.~\ref{fig:scatterplot_empirical}(a) and (d) for the ACF and PSD methods, respectively). We set bins along the $\alpha$ axis and plotted the average $\gamma_{\unbiased}$ within each bin. The average line (red) agrees with the theoretial line (black) well, strongly supporting the validity of the LMF prediction even at the quantitative level. We also provide a box plot in Fig.~\ref{fig:scatterplot_empirical}(b), where the statistical quantities, such as the first, second, and third quartiles, are calculated within each bin along the $\alpha$ axis. Furthermore, we provide the empirical PDF of the errors $\eta:=\alpha-1-\gamma_{\unbiased}$ as shown in Figs.~\ref{fig:scatterplot_empirical}(c) and (f) for the ACF and PSD methods, respectively. Since the average error is small $\la \eta^{\acf} \ra = 0.03$ for the ACF method and $\la \eta^{\psd}\ra= 0.003$ for the PSD method, respectively, our statistical analysis is self-consistent. 

		For reference, the original scatterplot between $\alpha$ and $\gamma_{\naive}$ (i.e., the consistent but biased estimator) is provided in Appendix~\ref{sec:app:scatterplot_NLLS}. In addition, we provide the three-yearly scatterplots between $\alpha$ and $\gamma_{\unbiased}$ as a robustness check (i.e., three-fold cross validation) in Appendix~\ref{sec:app:CrossValidation}.

		We have shown that the power-law exponent $\gamma$ in the ACF $C(\tau)$ is directly related to the microscopic power-law exponent $\alpha$ in the metaorder-length PDF $\rho(L)$. Since $\alpha$ is not observable from public data, our result implies that the LMF theory is useful for statistical estimation of microscopic parameters from public data. 

	\subsection{Discussion 1: estimation of the total number of traders}
		Since we show the feasibility of statistical estimation of $\alpha$ from the ACF power-law exponent $\gamma$, it is a natural idea to infer other microscopic quantities from the ACF prefactor $c_0$. In this subsection, we discuss the estimation of the total number of STs $N_{\ST}$ from the ACF prefactor $c_0$ based on the LMF theory. 

		\subsubsection{Review of the original LMF theory on the prefactor $c_0$}
			The LMF theory predicts that the ACF prefactor should be given by 
			\begin{equation}
				c_0^{\rm LMF} := \frac{1}{\alpha N_{\ST}^{2-\alpha}}
				\label{eq:LMF_prefactor_original}
			\end{equation}
			on the assumption that the intensity distribution $\{\lambda^{(i)}\}_{i\in \Omega_{\ST}}$ among the order-splitting traders is uniform, such that 
			\begin{equation}
				\lambda^{(i)} = \frac{1}{N_{\ST}} \>\>\> \mbox{for any }i \in \Omega_{\ST}.
				\label{eq:homogeneous_lambda_assumption}
			\end{equation}
			This prediction is applicable to infer the total number of the order-splitting traders $N_{\ST}$, such that 
			\begin{equation}
				N_{\ST}\approx N_{\ST}^{\rm LMF}(c_0,\gamma) := \left[\frac{1}{(\gamma+1)c_0}\right]^{\frac{1}{1-\gamma}},
				\label{eq:LMF_pred_N_ST}
			\end{equation}
			where the right-hand side is composed of publicly-available quantities from the sample ACF or PSD. Let us call $N_{\ST}^{\rm LMF}(c_0,\gamma)$ the LMF estimator for the total number of the STs. Since $N_{\ST}$ is not observable from public data, the prediction~\eqref{eq:LMF_pred_N_ST} is appealing from both academic and practical viewpoints. 

			Note that the LMF estimator has the singularity $\gamma=1$ at which the estimation fails in principle. Therefore, it is more realistic to study 
			\begin{equation}
				\left[N_{\ST}^{\rm LMF}(c_0,\gamma)\right]^{1-\gamma} = \frac{1}{(\gamma+1)c_0}
			\end{equation}
			by removing the singularity at $\gamma=1$.

		\subsubsection{Review of a generalised LMF theory on the prefactor $c_0$}
			During our data analysis, however, we noticed that the assumption~\eqref{eq:homogeneous_lambda_assumption} for homogeneous intensities is very unrealistic because time intervals between submissions broadly distributed in our dataset. Furthermore, in Ref.~\cite{SatoJSP2023}, the authors recently proposed a generalised LMF model by incorporating the inhomogeneous intensities and clarified the following points: 
			\begin{enumerate}
				\item Let us assume that all traders are order-splitting traders, but their intensity distribution is non-uniform, such that $\lambda^{(i)}\neq \frac{1}{N_{\ST}}$ for some $i \in \Omega_{\ST}$. 
				\item The ACF power-law exponent formula~\eqref{eq:LMF_prediction} robustly holds for any intensity distributions $\{\lambda^{(i)}\}_{i\in \Omega_{\ST}}$.
				\item On the other hand, the ACF prefactor formula~\eqref{eq:LMF_prefactor_original} is very sensittive to the system-specific details and does not hold anymore for general $\{\lambda^{(i)}\}_{i\in \Omega_{\ST}}$. Instead, the prefactor formula is replaced with 
				\begin{equation}
					c_0^{\rm SK} := \frac{1}{\alpha}\sum_{i\in \Omega_{\ST}}\left(\lambda^{(i)}\right)^{3-\alpha}.
				\end{equation}
				\item Furthermore, the homogeneous LMF formula~\eqref{eq:LMF_prefactor_original} systematically underestimates the actual prefactor, in the sense that 
				\begin{equation}
					c_0^{\rm SK} \geq c_0^{\rm LMF}.
				\end{equation}
			\end{enumerate}
			These heterogeneous LMF results imply that the LMF estimator $N_{\ST}^{\rm LMF}(c_0,\gamma)$ provides a lower bound of the true $N_{\ST}$:
			\begin{equation}
				N_{\ST}^{\rm LMF}(c_0,\gamma) \lesssim N_{\ST}.
				\label{ineq:heteroLMF_N_ST}
			\end{equation}

		\subsubsection{Scatterplot between $N_{\ST}^{\rm LMF}(c_0,\gamma)$ and $N_{\ST}$}
			
		\begin{figure*}
				\centering
				\includegraphics[width=170mm]{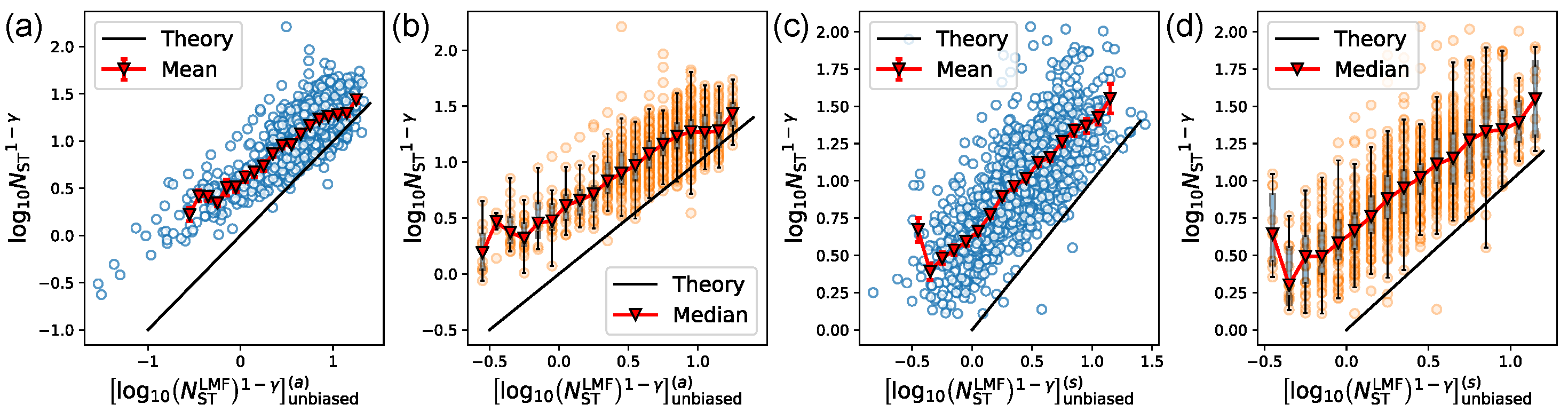}
				\caption{
					Estimation of the total number of the STs via the LMF theory for the ACF (a, b) and PSD (c, d) methods. (a, c)~Scatterplots between the unbiased LMF estimators $[\log_{10} \left(N_{\ST}^{\rm LMF}\right)^{1-\gamma_{\naive}}]_{\unbiased}$ and the actual values of $\log_{10}\left(N_{\ST}\right)^{1-\gamma_{\naive}}$. (b, d)~Corresponding boxplots. We find that the LMF estimator is strongly correlated with the actual $N_{\ST}$. However, the LMF estimator systematically underestimated the actual $N_{\ST}$, which is consistent with the theoretical inequaloty~\eqref{ineq:heteroLMF_N_ST} for a generalised LMF model in \cite{SatoJSP2023}. 
				}
				\label{fig:emp_estimate_N}
			\end{figure*}

			On the basis of the above theoretical predictions, we drew the scatterplot Fig.~\ref{fig:emp_estimate_N} between the true $\log_{10} \left(N_{\ST}\right)^{1-\gamma}$ and the LMF estimator $\log_{10} \left(N_{\ST}^{\rm LMF}\right)^{1-\gamma}$ after the finite-sample-size bias is removed (see Appendix~\ref{sec:app:N_ST_via_LMFprefactor} for the details). Since we should stick to empirically-available quantities, the estimators are based only on $\gamma_{\naive}$ and $c_{0,\naive}$ for both ACF and PSD methods. The ACF and PSD methods were employed (Fig.~(a, b) and (c, d), respectively), and they showed consistent results. We observed that the LMF estimator $N_{\ST}^{\rm LMF}$ is highly correlated with the true $N_{\ST}$. This fact implies that the ACF prefactor $c_0$ has potentially-useful information on $N_{\ST}$ in principle. On the other hand, the LMF estimator $N_{\ST}^{\rm LMF}$ systematically underestimates the actual value of $N_{\ST}$, which is consistent with the theoretical expectation~\eqref{ineq:heteroLMF_N_ST} for the heterogeneous LMF model. Therefore, the LMF estimator $N_{\ST}^{\rm LMF}$ should be interpreted as a lower bound of the total number of STs. 

			We thus conclude that the LMF theory qualitatively works even for the estimation of $N_{\ST}$ only from public data. However, its theoretical estimation is systematically biased due to traders' heterogeneity in order-splitting strategies. In this sense, strategies' heterogeneity needs to be considered for a more quantitative estimation of $N_{\ST}$.

	\subsection{Discussion 2: relation to previous results}
		In the original article~\cite{LMF}, Lillo, Mike, and Farmer showed a scatterplot between $\alpha$ and $\gamma$ by using the off-book market data as an imperfect proxy. While their figure does not statistically reject the LMF prediction~\eqref{eq:LMF_prediction} due to small sample size, it does not strongly support the validity of the strong LMF prediction~\eqref{eq:LMF_prediction} at the quantitative level. On the contrary, we have provided a much clear statistical evidence as Fig.~\ref{fig:scatterplot_empirical} with enough sample size, which strongly supports the validity of the LMF prediction, even at the quantitative level. Furthermore, we have successfully demonstrated that the ACF prefactor has the information on the total number of the order-splitting traders $N_{\ST}$ as shown in Fig.~\ref{fig:emp_estimate_N}.
		
		There are two more technical advantages of our statistical method than the previous one. The first advantage is that we directly estimated the metaorder-length (run-length) distribution at the level of individual traders, instead of the metaorder-volume distribution. This is in contrast to the statistical analysis in Ref.~\cite{LMF}, which is based on the metaorder-volume distribution in the absence of appropriate dataset. When one uses the metaorder-volume distribution to estimate $\alpha$, one has to assume that STs split their orders with constant volume for statistical analyses. However, this assumption is not realistic because volume specified by a market order is known to obey the power-law distributions empirically. On the other hand, we utilised a proper dataset and directly measured the metaorder-length distribution. This analysis does not require the assumption of the constant volume splitting. In this sense, we believe that our estimated exponent $\alpha$ would be more reliable for the calibration to the LMF model. 

		The second advantage is that we used the NLLS-based unbiased estimator $\gamma_{\unbiased}$, instead of the DFA estimator $\gamma_{\DFA}$. In econophysics, several researchers use the Hurst-exponent analysis based on the DFA to measure $\gamma$, and the scatterplot of Ref.~\cite{LMF} (see Fig.~\ref{fig:LMFPaper}) is also based on $\gamma_{\DFA}$. However, we find that $\gamma_{\DFA}$ has a serious problem in terms of the finite-sample-size bias at least in our numerical LMF simulations, and have concluded that $\gamma_{\DFA}$ is an inappropriate estimator in validating the LMF model. Since $\gamma_{\DFA}$ is not consistent with the LMF prediction~\eqref{eq:LMF_prediction}, we believe that our statistical estimation has a much greater advantage than the previous one.

	\subsection{Discussion 3: Implication for liquidity measurement}
		Our result strongly supports the validity of the LMF model even at the quantitative level~\eqref{eq:LMF_prediction}. Since the LMF model is based on the order-splitting hypothesis, we believe that our result is relevant to quantitative measurement of the market liquidity from a different angle. 

		According to the order-splitting hypothesis, traders split their large metaorders in the lack of revealed liquidity: the volumes at the best prices are too small compared with the metaorder volume, and traders have no choice but to split their orders into pieces. In the LMF model, the order book for markets with smaller $\alpha$ and large $N_{\ST}$ is not thick enough for many institutional investors to immediately execute metaorders. Thus, the parameter set $(\alpha, N_{\ST})$ characterises the illiquidity of markets regarding metaorder splittings. Particularly, $N_{\ST}$ characterises how many institutional investors are waiting for the order books to replenish during their order splitting, which might have signifinant meaning for platform managers.	Such an aspect of liquidity shortage is not measured by traditional liquidity measures, such as market spread and market impact. We believe that it might be interesting to develop some liqidity measures based on the order-splitting hypothesis as an another direction.

		\subsection{Discussion 4: Open questions on statistical analyses}
		We approximately measured the power-law exponent of the sign ACF by the NLLS-based unbiased estimator $\gamma_{\unbiased}$. While we believe that this estimator is practically reliable at least for our dataset, we are not sure whether this is always the best option. Indeed, the approximate construction of the unbiased estimator is based on the numerical observation of the approximate linear regression relation~\eqref{eq:linear_regression} for $\alpha-1 \in (0,1)$, which is not theoretically proved yet. In addition, this construction of unbiased estimators may depend on selection of underlying microscopic models (i.e., the LMF model in our case). Seeking the optimal unbiased estimator is an urgent topic as statistics. 

		In addition, we found a serious problem of the DFA estimator $\gamma_{\DFA}$ in terms of the finite-sample-size bias. Since we are not sure about its critical reason, this problem should be sought more deeply from the viewpoint of statistical analyses. In particular, we are interested in its robustness in terms of the consistency: i.e., does $\gamma_{\DFA}$ coincide with the true $\gamma$ for the infinite sample size, even if the time series is generated by some microscopic model, instead of the fractional Brownian motion? Anyway, the long-memory character of the LRC is a huge obstacle for statistical analyses, and thus development of statistical methods will be important.

\section{Concluding remarks}\label{sec:conclusion}
	While the LMF model has been a cornerstone to support the order-splitting hypothsis, its prediction~\eqref{eq:LMF_prediction} has not been verified at the quantitative level. In this paper, we have quantitatively established the validity of the LMF prediction~\eqref{eq:LMF_prediction} by analysing a large dataset of the TSE market over nine years. We first identified the RTs and STs by clustering analysis, and measured the microscopic power-law exponent $\alpha$ in terms of the metaorder length for STs. We then develop a statistical method to measure the power-law exponent $\gamma$ in the sign ACF. The scatterplot between $\alpha$ and $\gamma$ is provided as the main result, strongly supporting the validity of the LMF prediction~\eqref{eq:LMF_prediction}. Furthermore, we discuss a practical method to estimate the total number of order splitters from the ACF prefactor on the basis of the LMF theory.

	Our study builds upon the stream of ecological analyses of financial markets, which is based on the trading-strategy clustering at the level of individual traders. In the literature, one of the pioneering studies on trading-strategy clustering was provided in Ref.~\cite{MantegnaNJP2012} in 2012 by focusing on market-order submissions. As for limit-order submissions, strategy clustering was first provided by Ref.~\cite{KanazawaPRL,KanazawaPRE,Sueshige2018,Sueshige2019} for the EBS FX market in 2018 (i.e., regarding trend-following behaviour), and was also provided by Ref.~\cite{Goshima2019} for the TSE market in 2019 (i.e., regarding market-making bahaviour). In this work, we classify traders into RTs and STs in terms of market-order submissions. It will be interesting to investigate the roles of RTs and STs in the ecology of the TSE market. We believe that this research direction would be promising in developing market microstructure for the future.

\begin{acknowledgments}
	YS was supported by JST SPRING (Grant Number JPMJSP2110).
	KK was supported by JST PRESTO (Grant Number JPMJPR20M2), JSPS KAKENHI (Grant Numbers 21H01560 and 22H01141), and JSPS Core-to-Core Program (Grant Number JPJSCCA20200001). We greatly appreciate the data proivision and careful review of this report by the JPX Group, Inc.

	Here we describe the author contribution to this study. YS contributed the numerical and empirical analyses by program coding. KK designed the research plot and supervised the project. YS and KK wrote the manuscript and agree with all the findings. 

	We declare no financial conflict of interest. The JPX Group, Inc. provided the original data for this study without any financial support.
\end{acknowledgments}

\appendix
\section{Data availability}\label{sec:app:dataAvailability}
	The data that support the findings of this study was provided from the JPX Group, Inc. However, restrictions apply to the availability of these data, which were used under license only for our projects. Therefore, the authors are not allowed to distribute the data without the explicit permission of the JPX Group, Inc. 

\section{Intraday seasonality}\label{sec:app:IntradaySeasonality}
	In our data analysis, we excluded the data during the periods around the opening and closing auctions. This exceptional rule is applied to avoid the intraday-seasonality effect, which is a stylised fact in various financial markets~\cite{BouchaudText}. In this appendix, we show the statistical evidence of the intraday seasonality in TSE, called the U-shape profile of the temporal market-order activity, to justify our exceptional rule. 

	In this appendix, we use the physical time $t$ (minutes), representing the elapsed time from the starting time of the morning continuous double auction (9:00 JST) with the lunch break (11:30-12:30 JST) excluded. For example, $t=0$ represents 9:00 JST, $t=150$ represents 11:30 JST, $t=151$ represents 12:31 JST, and $t=300$ represents 15:00 JST.

\begin{figure}
		\centering
		\includegraphics[width=150mm]{"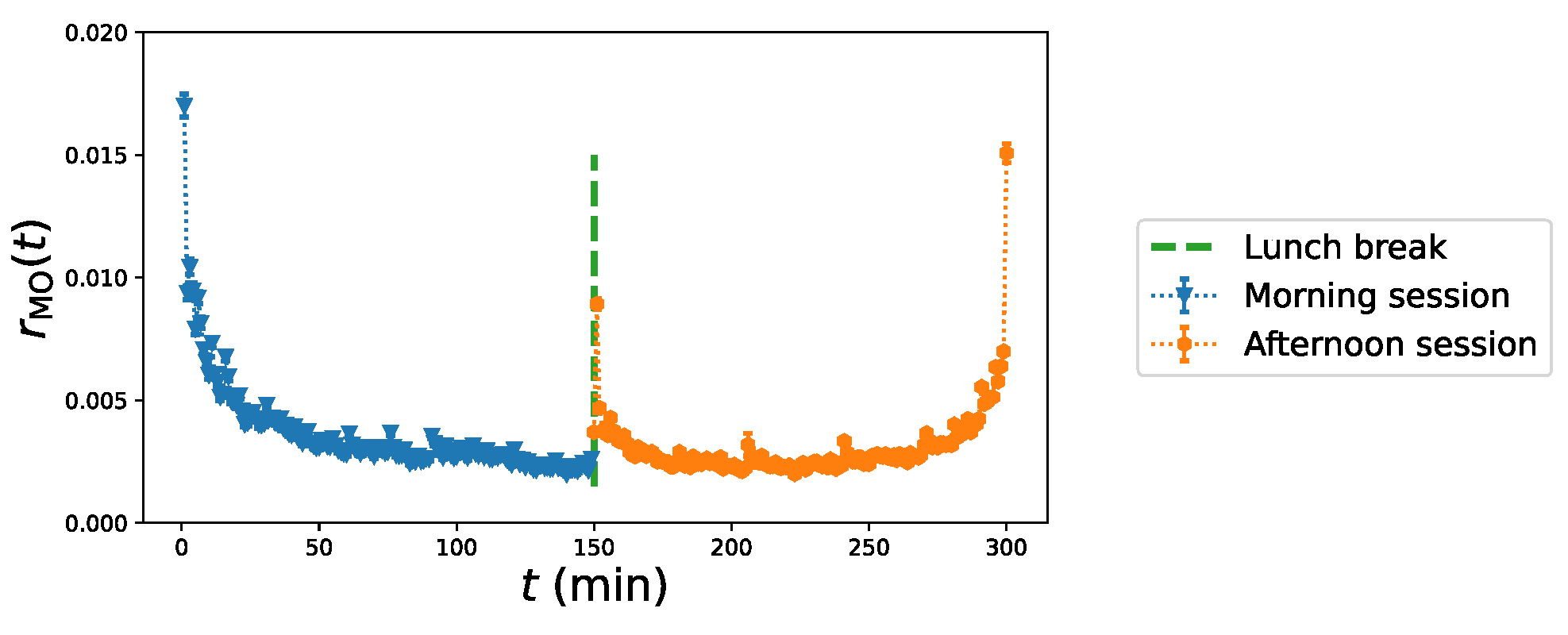"}
		\caption{
				Temporal market-order activity of Toyota 2020. The yearly average of the temporal market-order ratio $r_{\rm MO}$ is plotted to show the U-shape profile in its intensity: i.e., the transactions are active near the opening and closing times of the continuous double auction $t=0$, $t=150$, and $t=300$. Note that $r_{\rm MO}(t)$ is defined by $r_{\rm MO}(t):=N_{\rm MO}(t)/N_{\rm MO}^{\rm tot}$ with the daily-total number of market orders $N_{\rm MO}^{\rm tot}$ and minutely-total number of market orders $N_{\rm MO}(t)$. 
		}
		\label{fig:MO}
	\end{figure}

	Let us focus on Toyota Motor Corporation in 2020. The daily-total number of market orders is written as $N_{\rm MO}^{\rm tot}$ and the number of market orders during $[t,t+1)$ is written as $N_{\rm MO}(t)$. The temporal market-order ratio is then defined by $r_{\rm MO}(t):=N_{\rm MO}(t)/N_{\rm MO}^{\rm tot}$. In Fig.~\ref{fig:MO}, we plotted the yearly average of the temporal market-order ratio $r_{\rm MO}(t)$. This figure shows the market-order submissions are active around the opening and closing times of the continuous double auctions (i.e., $t=0$, $t=150$, and $t=300$), consistently with the empirical ``U-shape profiles" in previous reports~\cite{BouchaudText}.

\section{Numerical implementation of the random integer number obeying a power law}
\label{app:random_integer_powerlaw}
	Here we describe the numerical method to generate random integer numbers obeying a power-law relation 
	\begin{equation}
		P(L) \propto L^{-\alpha-1} \>\>\> \mbox{for large }L
	\end{equation}
	with an exponent $\alpha > 1$. Let us consider a continuous positive random number $x \in [1,\infty)$, which obeys the continuous Pareto distribution
	\begin{equation}
		P(x) =\alpha x^{-\alpha-1} \>\>\> (x \in [1,\infty)).
	\end{equation}
	The pareto random number $x$ can be generated by 
	\begin{equation}
		x := \frac{1}{(1-u)^{1/\alpha}}
	\end{equation}
	with a uniform random number $u\in [0,1)$. Finally, the integer random number $L$ is given by 
	\begin{equation}
		L := \lfloor x \rfloor,
	\end{equation}
	where the floor function $\lfloor x \rfloor$ signifies the maximum integer not larger than than $x$. 

\section{Summary of the technical problems to be solved}\label{sec:app:summary_technicalproblems}
	The validation steps for the LMF prediction is rather straightforward. Why this relationship~\eqref{eq:review_LMP_quantPred} has not been verified yet? In our view, there are three technical problems in proving the quantitaive prediction~\eqref{eq:review_LMP_quantPred}. Let us briefly summarise these technical problems one by one. 
	
	The first problem would be the scarcity of necessary high-quality data. To measure $\alpha$, we are required to identify order-splitting traders (STs) at the level of individual traders by applying strategy clustering, and then measure the empirical PDF of the runs as $P(L)$. For example, let us write the set of STs as $\Omega_{\ST}$. The empirical PDF of the runs are obtained 
	\begin{equation}
		P(L) \propto \sum_{i\in \Omega_{\ST}}\sum_{k}\delta(L-L_{k}^{(i)}).
	\end{equation}
	Since the STs $\Omega_{\ST}$ and their run sequences $\{L_{k}^{(i)}\}_{k,i\in \Omega_{\ST}}$ are the necessary inputs for the run PDF $P(L)$, we have to analyse the datasets enabling us to track orders at the level of trader accounts. However, such high-quality data are very scarce in terms of the data availability. 
	
	The second problem would be the necessary datasize. While there are a few studies analysing account-level datasets, the necessary datasize would be expected huge in verifying Eq.~\eqref{eq:review_LMP_quantPred}. Indeed, the inputs of the scatterplot are the power-law exponents $\alpha$ and $\gamma$, and their accurate measurements are not easy: theoretically, they are expected to distribute typically within the range $\alpha \in (1,2)$ and $\gamma \in (0,1)$. Therefore, it would be necessary to control their estimation errors roughly less than $0.1$. In particular, the accurate estimation of $\gamma$ is very hard. Assuming that the datapoints of $C(\tau)$ are necessary to cover the range $\tau \in (10^1, 10^3)$, we have to suppress the noise in the ACF at a low level even at $\tau\sim 10^3$. Through our numerical simulations of the LMF model, we estimate that a long order-sign sequence $\{\eps(t)\}_t$ is necessary, such as $N_{\eps}>5\times 10^5$ at least, even to obtain one datapoint $(\alpha,\gamma)$ in the scatterplot.

	The third problem is related to the fact that the LMF model belongs to the long-memory process (see Chapter 10 in Ref.~\cite{BouchaudText}), implying that the convergence speed of any sample mean is slower than usual in terms of the sample size $N_{\eps}$. Indeed, the long-memory character $C(\tau)\approx \tau^{-\gamma}$ with $\gamma \in (0,1)$ suggests that $\eps(t)$ and $\eps(t+\tau)$ are not statistically independent with each other even for large $\tau>0$. Thus, the estimation of $\gamma$ will require a large dataset from such a theoretical viewpoint.

	To overcome such technical difficulties, in this report, we analyse a large TSE dataset provided by the JPX Group, Inc. This data not only includes the account-level information (i.e., the virtual-server IDs), but also covers all the stocks over nine years. We then finally report the first verification of the quantitative LMF prediction~\eqref{eq:review_LMP_quantPred} from the viewpoint of the big-data analysis.

\section{Metaorder-length distribution for RTs}\label{app:metaorder_RTs}
	\begin{figure}
		\centering
		\includegraphics[width=150mm]{"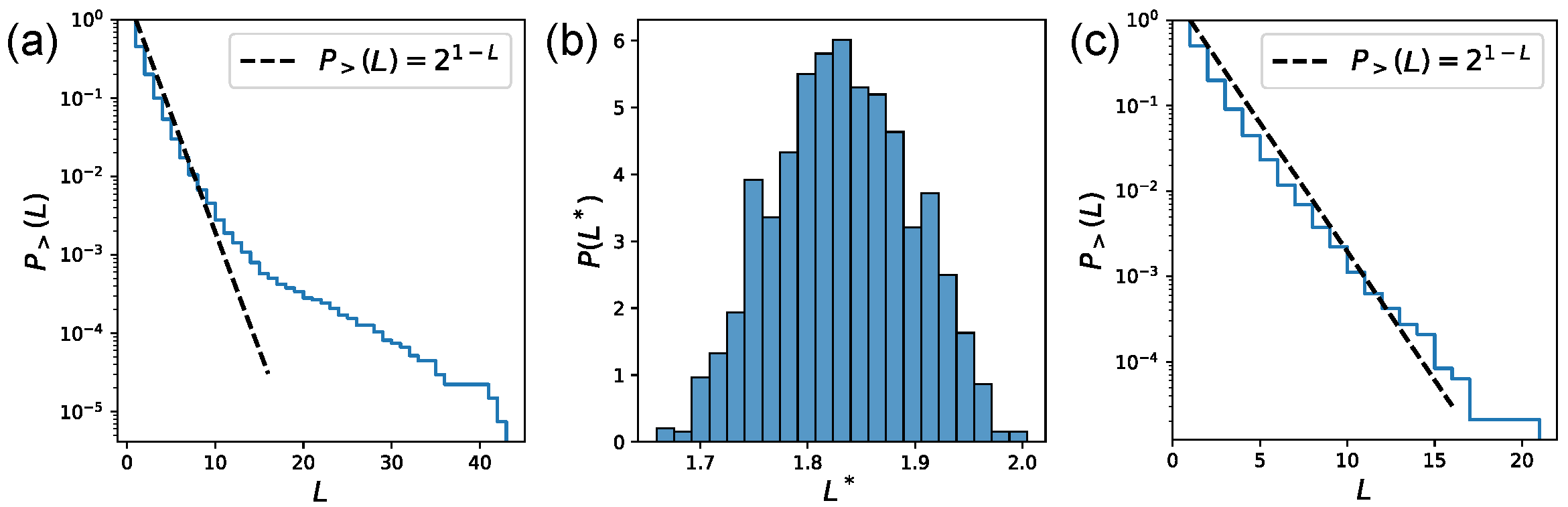"}
		\caption{
				Characters of metaorder-length distribution on RTs.
				(a)~The metaorder-length (run-length) CCDF of the all RTs for Toyota Motor Corporation in 2020. The run-length CCDF obeys the exponential law at least for the body part $L\lesssim 10$, such that $P_>(L)\sim 2^{1-L}$, as theoretically expected. At the same time, we observed a theoretical discrepancy for the tail because the second-kind statistical error was not controlled in our binomial test; a small portion of STs might be included in the RT cluster.
				(b)~The empirical PDF $P(L^*)$ of the decay length $L^*$ in our dataset, by assuming the exponential metaorder-length CCDF $P_>(L)\approx (L^*)^{1-L}$ for the RTs for each data point. The decay length $L^*$ is measured by the maximum likelihood estimation (i.e., $L^* = \la L^{\RT}\ra$). The PDF has a sharp peak around $L^*\approx 1.8$, consistent with the theoretical prediction~\eqref{eq:dist_random_theory}. 
				(c)~The aggregated metaorder CCDF only for the active RTs (submitting more than one thousand orders annually), showing the exponential law without fat tails for Toyota 2020. This is an empirically successful symptom of reducing the second kind's error.
		}
		\label{fig:agg_meta_RTs}
	\end{figure}
	In Sec.~\ref{sec:binomial_test}, we regard any trader as an RT when the binomial test was not rejected with the significance level $\theta=0.01$. In the standard theory of statistical tests, it is often emphasised that passing tests does not necessarily mean the acceptance (proof) of the null hypothesis, and we should not draw hasty conclusions: the error of the first kind (false-positive rate) is controlled within the significance level $\theta$ for the rejection, but the error of the second kind (false-negative rate) is not controlled for the ``acceptance" in the statistical tests. In this sense, while our clustering method is expected to be reasonable in identifying the set of STs within the significance level, the identification of the set of RTs might be incomplete; some small part of non-RTs, such as STs, might be included even in the RT cluster since we did not control the error of the second kind. 

	While we acknowledge this possible incompleteness of our clustering method for RTs, it would be helpful to check whether the set of RTs satisfies our theoretical expectation for reference. If the assumption of the null hypothesis (i.e. the symmetric Bernoulli process) is exactly correct, the CCDF of the run lengths for RTs should be given by the exponential distribution: 
	\begin{equation}
		P_>(L) = 2^{1-L} \label{eq:dist_random_theory}
	\end{equation} 
	for any positive integer $L$. We check this character regarding the RT clusters. 

	We studied the metaorder-length distribution for the RTs: we consider the joint run-length sequences for RTs and the corresponding empirical metaorder-length CCDF: 
	\begin{equation}
		\left\{L^{\RT}_k\right\}_{k} := 
		\bigcup_{i \in \Omega_{\RT}} \left\{L_k^{(i)}\right\}_{k}, \>\>\> 
		P_>\left(L^{\RT}\right) := \frac{N_>(L^{\RT})}{\left|\{L^{\RT}_k\}_{k}\right|}, \>\>\> 
		N_>(L^{\RT}) := \int_{L^{\RT}}^\infty dy\sum_{k} \delta\left(y-L^{\RT}_k\right).
	\end{equation}

	The empirical metaorder CCDF for the RTs is plotted in Fig.~\ref{fig:agg_meta_RTs}(a) for Toyota Motor Corporation in 2020. This plot shows that the metaorder CCDF exhibits the exponential law $P_>(L)\approx 2^{1-L}$ for the body part $L\lesssim 10$. In addition, the estimated decay length $L^*$ shows a sharp peak around $L^*\approx 1.8$ based on the maximum likelihood estimation $L^*=\langle L^{\RT}\rangle$ for the exponential law $P_>(L^{\RT})\approx (L^*)^{1-L^{\RT}}$ (see Fig.~\ref{fig:agg_meta_RTs}(b)). This result shows a minimum self-consistency of our clustering algorithm with $L^*\approx 2.0$. At the same time, we observe the discrepancy from the exponential law for the tail part $L\gtrsim 10$. This discrepancy is reasonable because we did not control the statistical error of the second kind, and a small portion of STs might be included in the RT cluster. 

	Improvement of our clustering algorithm is a future open issue regarding the RTs, and applying some filters would be desirable to control the second kind's error. As an initial attempt, we applied a simple filter by focusing only on active RTs submitting more than one thousand market orders an year (i.e., a few submissions everyday on average). We considered this filter a reasonable candidate, because a small portion of inactive RTs seemigly submitted large metaorders only a few times during the year while they behaved as RTs during most of the time. The aggregated metaorder CCDF only for the active RTs are plotted in Fig~\ref{fig:agg_meta_RTs}(c) for Toyota 2020, where the discrepancy at the tail disappears. We checked all the stocks in 2012 and 2020 by eyes and found a similar observations.

\section{Detailed implementation of the nonlinear least squares}\label{sec:app:NLLS}
	In this Appendix, we describe the measurement of the power-law exponent $\gamma$ and the prefactor $c_0$ in the sign ACF $C(\tau)\simeq c_0\tau^{-\gamma}$ for large $\tau$. Our methods are based on the nonlinear least squares (NLLS) for the ACF and PSD.

\subsection{NLLS estimators based on the sample ACF}\label{sec:app:NLLS_ACF}
	\begin{figure}
		\centering
		\includegraphics[width=150mm]{"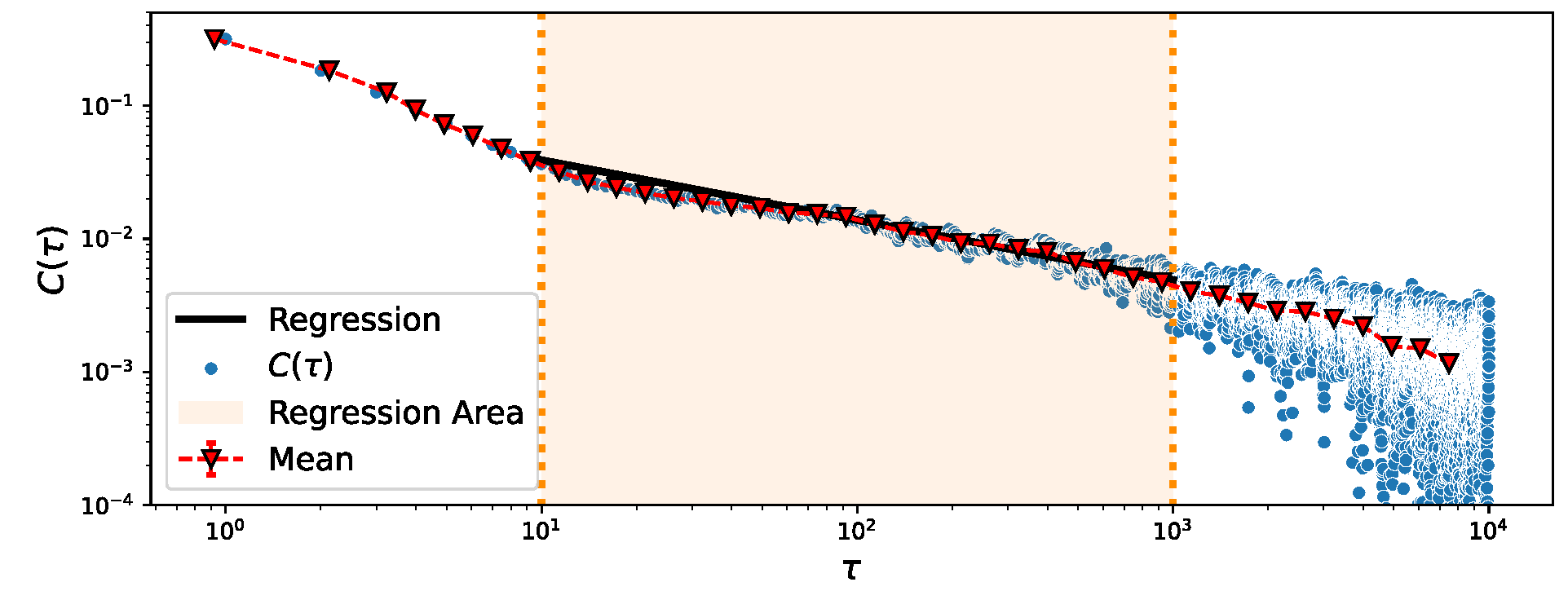"}
		\caption{
				Schematic figure of the ACF fitting based on the NLLS estimation for Toyota 2020. We used the orange area $\tau\in [\tau_{\threshold}^{-},\tau_{\threshold}^{+}]$ for the fitting to obtain the power-law guideline with exponent $\gamma_{\naive}$ (solid black).
		}
		\label{fig:sample_fitting}
	\end{figure}
	We first describe the NLLS estimation for the empirical ACFs. The basic idea is to systematically fix the fitting range $[\tau_{\threshold}^-, \tau_{\threshold}^+]$, and then apply the power-law fitting $C(\tau)\propto \tau^{-\gamma_{\naive}^\acf}$ to the sample ACF for the range $[\tau_{\threshold}^-, \tau_{\threshold}^+]$ (see Fig.~\ref{fig:sample_fitting} for the scheme). The detailed process is given as follows: 
	
	\subsubsection*{Step 1: the sample ACF}
		The sample ACF is defined by 
		\begin{equation}
			C_{\sample} (\tau) := \frac{1}{N_{\eps}-\tau}\sum_{t=1}^{N_{\eps}-\tau} \eps(t)\eps(t+\tau)
		\end{equation}
		for positive $\tau>0$, by assuming the symmetry $\la \eps(t)\ra=0$ for the range $\tau \in [1,10^4]$. This symmetric assumption is commonly used in other literature~\cite{Toth2015} and its validity is also checked in our dataset. For non-positive $\tau$, we define $C(\tau)=0$. 
		
	\subsubsection*{Step 2: the lower threshold}
		As reported in various datasets~\cite{BouchaudText}, the sample ACF initially exhibits a relatively rapid decay for small $\tau$ and the power-law decay follows for large $\tau$ in our datasets. To estimate the power-law exponent, it will be useful to estimate the lower bound $\tau_{\threshold}^-$ for the final fitting regime. This threshold is estimated as follows: let us first estimate the initial decay timescale $\tau_{\temp}$ by using the NLLS fitting of the sample ACF with tentative fitting function
		\begin{align}
			C_{\text{model}}(\tau) = C_{\temp}^{(0)} e^{-\tau/\tau_{\temp}} + C_{\temp}^{(1)} \tau^{-\gamma^{(1)}_{\temp}}\label{eq:fit1}
		\end{align}
		with the temporary fitting parameters $C_{\temp}^{(0)}$, $C_{\temp}^{(1)}$, $\tau_{\temp}$, and $\gamma^{(1)}_{\temp}$ for the range $\tau \in [1,10^3]$. These parameters are estimated by the relative least squared error (RLS) method (see Appendix~\ref{sec:app:RLS_method}).

		Based of the tentative fitting formula~\eqref{eq:fit1}, we next fix the lower threshold $\tau_{\threshold}^-$ between the exponential and power-law decays as follows: since we would like to estimate the lower threshold $\tau_{\threshold}^-$ that satisfies $|C_{\temp}^{(0)}e^{-\tau_{\threshold}^-/\tau_{\temp}}| \ll |C_{\temp}^{(1)}\left(\tau_{\threshold}^-\right)^{-\gamma^{(1)}_{\temp}}|$, let us consider the area where the power-law part is dominant:
		\begin{equation}
			A_{\rm pow} := \left\{\tau    \,\,\Bigg|\,\, 
			\bigg|\frac{C_{\temp}^{(0)}e^{-\tau/\tau_{\temp}}}{C_{\temp}^{(1)}\tau^{-\gamma^{(1)}_{\temp}}}\bigg| < \epsilon_{\threshold},\>\> 10\leq \tau\leq 10^2  \right\}
		\end{equation}
		with a small parameter $\epsilon_{\threshold}:= 0.1$. The lower threshold $\tau_{\threshold}^-$ is defined by 
		\begin{equation}
			\tau_{\threshold}^{-} := 
			\begin{cases}
				\min_{A_{\rm pow}} \tau & \mbox{if $A_{\rm pow}$ is not empty} \\
				10^2 & \mbox{if $A_{\rm pow}$ is empty}
			\end{cases}.
		\end{equation}

	\subsubsection*{Step 3: logarithmic smoothing}
		The sample ACF exhibits fluctuations, particularly for large $\tau$, due to the finite sample size. To remove such statistical fluctuations, we define the smoothed sample ACF:
		\begin{subequations}\label{eq:ACF_smooth_log}
		\begin{align}
			C_{\smooth}(\tau) := \sum_{t=-\infty}^\infty w_{\delta}(\tau;t)C_{\sample}(t), \>\>\> 
			w_{\delta}(\tau;t) := \begin{cases}
								\displaystyle 
								\frac{1}{\tau_{\smooth}^+(t)-\tau_{\smooth}^-(t)} & (t \in \left( \tau_{\smooth}^-(\tau), \tau_{\smooth}^+(\tau) \right) \\
								0 & (t \not \in \left( \tau_{\smooth}^-(\tau), \tau_{\smooth}^+(\tau) \right)
							\end{cases}.
		\end{align}
		Since we are interested in the estimation of the power-law exponent, we use the logarighmic smoothing based on 
		\begin{equation}
			\tau_{\smooth}^+(\tau) = \left\lfloor \tau \>10^{+\delta/2}\right\rfloor, \>\>\>
			\tau_{\smooth}^-(\tau) = \left\lfloor \tau \>10^{-\delta/2}\right\rfloor
		\end{equation}				
		\end{subequations}
		with the smoothing window size $\delta=0.05$. This smoothing method is a discrete-time version of logarithmic smoothing for continuous time (see Appendix~\ref{sec:app:logarithmic_smoothing}).

	\subsubsection*{Step 4: the upper threshold}
		It would be desirable to observe the power-law decay in the region of about two digits on the log-log ACF plot, such as by setting $\tau_{\threshold}^{+} = 10^2 \tau_{\threshold}^{-}$. However, the sample ACF will be statistically insignificant for very large $\tau$ and such a naive setting of $\tau_{\threshold}^{+}$ might be inappropriate in general. 
		
		Indeed, even if the order-sign sequence were generated by the completely random manner (i.e., the white noise), the sample ACF can take non-zero values, such that $|C(\tau)| \simeq N_\epsilon^{-1/2}$, due to statistical errors. In this sense, if the absolute values of the sample ACF are smaller than $N_\epsilon^{-1/2}$, it is reasonable that the values of the sample ACF are regarded as statistically insignificant. 
		
		Based on this idea, we estimate an upper cutoff $\tau_{\stat}^{+}$ in terms of the statistical significance. The area where the smoothed ACF is statistically significant is estimated by 
		\begin{equation}
			A_{\stat} := \left\{\tau    \,\,\bigg|\,\, C_{\smooth}(\tau) < \frac{1}{\sqrt{N_\eps}} \>,\>  10^3 \leq \tau \right\}.
		\end{equation}
		The upper cutoff $\tau_{\stat}^{+}$ for statistical significance is estimated as 
		\begin{equation}
			\tau_{\stat}^{+} := 
			\begin{cases}
				\min_{A_{\stat}} \tau & \mbox{if $A_{\stat}$ is not empty} \\
				10^4 & \mbox{if $A_{\stat}$ is empty}
			\end{cases}.
		\end{equation}
		Finally, the upper threshold for our final fitting $\tau_{\threshold}^{+}$ is given by 
		\begin{equation}
			\tau_{\threshold}^{+} := \min \left\{\tau_{\stat}^{+},10^2\tau_{\threshold}^{-} \right\}.
		\end{equation}

	\subsubsection*{Step 5: the determination of $\gamma_{\naive}^\acf$}
		The power-law exponent $\gamma_{\naive}^\acf$ is finally estimated by the RLS fitting of the smoothed ACF $C_{\smooth}(\tau)$ for the range $[\tau_{\threshold}^-,\tau_{\threshold}^+]$ by the power-law fitting function
		\begin{align}
			C(\tau) \propto \tau^{-\gamma_{\naive}^\acf} .\label{eq:fit2}
		\end{align}
		with fitting parameters $\gamma_{\naive}^\acf$. 

	\subsubsection*{Step 6: the determination of $c_{0,\naive}^\acf$}
		Finally, we determine $c_{0,\naive}^\acf$ by integration\footnote{ The dimension of $c_{0,\naive}^\acf$ is given by [time$^{-\gamma}$], which is automatically consistent with the dimension analysis in this integration method. In addition, the integration of the ACF has a global smoothing effect, by which we expect that the estimation is more stable.} of the smoothed ACF $C_{\smooth}(\tau)$ as 
		\begin{align}
			c_{0,\naive}^\acf :=\frac{(1-\gamma_\naive^\acf)}{
				\left(\tau^{+}_\threshold \right)^{1-\gamma_\naive^\acf} 
				- \left(\tau^{-}_\threshold\right)^{1-\gamma_\naive^\acf}} 
				\int_{\tau^{-}_\threshold}^{\tau^{+}_\threshold} C_{\smooth}(\tau')d\tau'.
		\end{align}

\subsection{NLLS estimation based on the sample PSD}\label{sec:app:NLLS_PSD}
	\begin{figure*}
		\centering
		\includegraphics[width=150mm]{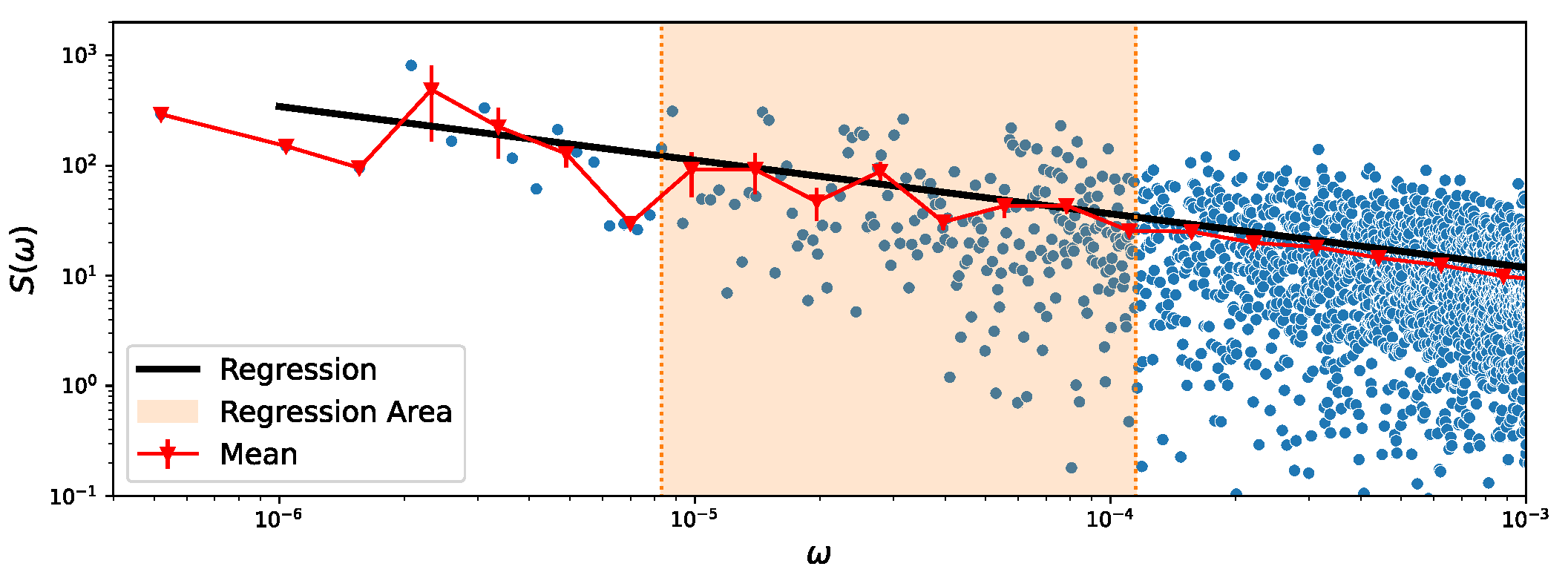}
		\caption{
			Schematic figure of the PSD fitting based on the NLLS estimation for Toyota 2020. We used the orange area $\tau\in [\omega_{\threshold}^{-},\omega_{\threshold}^{+}]$ for the fitting to obtain the power-law guideline with exponent $H_{\naive}$ (solid black).
		}
		\label{fig:sec:PSD_fitting}
	\end{figure*}
	We next describe the measurements based on the sample PSD. According to \cite{Tauber}, for $\gamma \in (0,1)$ (or equivalently $\alpha \in (1,2)$), the theoretical PSD of the LMF model is given by
	\begin{align}
		S(\omega)
		&\approx  \int^{\infty}_{0} d\tau e^{2 i \pi  \omega \tau }\left(\frac{N_{\rm ST}^{\alpha-2}}{\alpha}|\tau|^{-\gamma}\right)\\
		&= c^\spectra \omega^{\gamma-1},\>\>\>
		c^\spectra=\frac{N_{\rm ST}^{\alpha-2}}{\alpha}2^{\gamma}\pi^{\gamma-1} \Gamma(1-\gamma)\sin\left(\frac{\pi \gamma}{2}\right).\\
		&\sim \omega^{-H}, \>\>\> H =1-\gamma= 2-\alpha.\>\>\>\text{for small $\omega$},\label{eq:FourierAC}
	\end{align}
	where the Wiener-Khinchin theorem~\eqref{eq:WK-theorem} is used. Similarly to the ACF method, we apply the power-law fitting $S(\omega)\propto \omega^{-H_{\naive}}$ to the sample PSD for the range $[\omega_{\threshold}^-, \omega_{\threshold}^+]$. The fitting range is automatically fixed as follows.

	\subsubsection*{Step 1: the sample PSD}
		The sample PSD $S_{\sample}(\omega)$ was estimated by the periodogram method using {\it scipy}~\cite{Scipy}. 

		\subsubsection*{Step2: linear smoothing of the PSD}
		The sample PSD fluctuates due to the finite sample size. We apply normal smoothing of the empirical PSD\footnote{We used normal smoothing instead of logarithmic smoothing because we are interested in the low-frequency regime of the PSD.}:
		\begin{subequations}\label{eq:PSD_smooth_log}
			\begin{align}
				S_{\smooth}(\omega) := \sum_{\omega=-\infty}^\infty w_{\delta}(\omega)C_{\sample}(\omega), \>\>\> 
				w_{\delta}(\omega) := \begin{cases}
									\displaystyle 
									\frac{1}{2\delta+1} & (\omega \in \left[ f-\delta\Delta_\omega, \omega +\delta\Delta_\omega, \right] \\
									0 & (\omega \not \in  \left[ \omega-\delta\Delta_\omega, \omega +\delta\Delta_\omega, \right] 
								\end{cases}, \>\>\> 
				\Delta_\omega = \frac{1}{N_{\eps}}
			\end{align}
		\end{subequations}
		with the smoothing window size $\delta=5$.

	\subsubsection*{Step3: the lower and upper thresholds}
		Let us determine the lower and upper thresholds $\omega_{\threshold}^-$ and $\omega_{\threshold}^-$. First, we describe the method to fix the lower threshold $\omega_{\threshold}^-$. The smoothed PSD $S_{\smooth}(\omega)$ fluctuates near the lowest frequency $\omega\approx \Delta_{\omega}$, and we discarded some of the low-frequency datapoints. We set $\omega_{\threshold}^-=15\Delta_{\omega}$. 

		We next fix the upper threshold $\omega_{\threshold}^{+}$. According to the original LFM theory~\cite{Lillo2004}, the asymptotical relationships~\eqref{eq:FourierAC} is valid up to $\tau\gg |\Omega_\TR|$ (or equivalently $2\pi\omega \ll |\Omega_\TR|^{-1}$). In our dataset, the typical number of the splitting traders was $10^2$ (see Sec.~\ref{sec:ResultClassification}). We therefore assume that $\omega_{\threshold}^+$ should be set smaller than $10^{-3}$. 
		
		In addition, the PSD fluctuates for large frequency $\omega$ due to the finite-sample size. Let us define the half bandwidth $\omega_{\rm half}$ as the characteristic decay frequency of the PSD as: 
		\begin{equation}
			\omega_{\rm half} := \min_{B} \omega, \>\>\> 
			B := \left\{ 
				\omega \> \big| \> S_{\smooth}(\omega) < S_{\rm median},\>\>\> 
				10^2\Delta_\omega \leq \omega \leq 10^3\Delta_\omega 
				\right\}
		\end{equation}
		if $B$ is not empty, where $S_{\rm median}$ is the median of $\{S_{\smooth}(\omega)\}_{\omega\in[\omega_{\threshold}^-,10^{-3}]}$. Considering the possibility that $B$ might be empty in general, we set the upper threshold as 
		\begin{equation}
			\omega^{+}_\threshold:= 
			\begin{cases}
				\omega_{\rm half} & \mbox{if $B$ is not empty} \\
				10^3 \Delta_\omega & \mbox{if $B$ is empty}
			\end{cases}.
		\end{equation}

	\subsubsection*{Step 4: the determination of $\gamma_\naive^\spectra$}
		The Hurst exponent was estimated by the RLS fitting (see Appendix~\ref{sec:app:RLS_method}) to the smoothed PSD $S_{\smooth}(\omega)$ for the range $[\omega^{-}_{\threshold},\omega^{+}_{\threshold}]$, such that 
		\begin{align}
			S(\omega) \propto \omega^{-H_\naive}
		\end{align}
		with fitting parameter $H_\naive$. The NLLS power-law exponent $\gamma_{\naive}^{\psd}$ is measured by the PSD method using the asymptotic relationship~\eqref{eq:FourierAC}:
		\begin{align}
			\gamma_\naive^\spectra &:= 1-H_\naive\label{eq:GammaFourier}
		\end{align}
		if $H_\naive < 1$. 

		Due to methodological artefacts, we sometimes obtain $H_\naive > 1$, implying negative $\gamma$ (i.e., a monotonically increasing ACF, which does not make sense). Since this is an obvious symptom of estimation failure, we excluded such datapoints as exceptional handling.

	\subsubsection*{Step 5: the determination of $C_\naive^\spectra$}
		Finally, the ACF prefactor $c_{0,\naive}^\spectra$ is determined by the integration of the PSD, such that 
		\begin{align}
			c_{0,\naive}^\spectra :=
			\frac{2}{(2\pi)^{\gamma_\naive^\spectra-1} \Gamma(1-\gamma_\naive^\spectra)\sin\left(\frac{\pi\gamma_\naive^\spectra}{2}\right)}
			\frac{ \gamma_\naive^\spectra }{
				\left(\omega^{+}_\threshold \right)^{\gamma_\naive^\spectra} 
				- \left(\omega^{-}_\threshold\right)^{\gamma_\naive^\spectra}} 
				\int_{\omega^{-}_\threshold}^{\omega^{+}_\threshold}du S(u).
		\end{align}

\section{The scatterplot based on the NLLS estimator}\label{sec:app:scatterplot_NLLS}
	\begin{figure}
		\centering
		\includegraphics[width=175mm]{"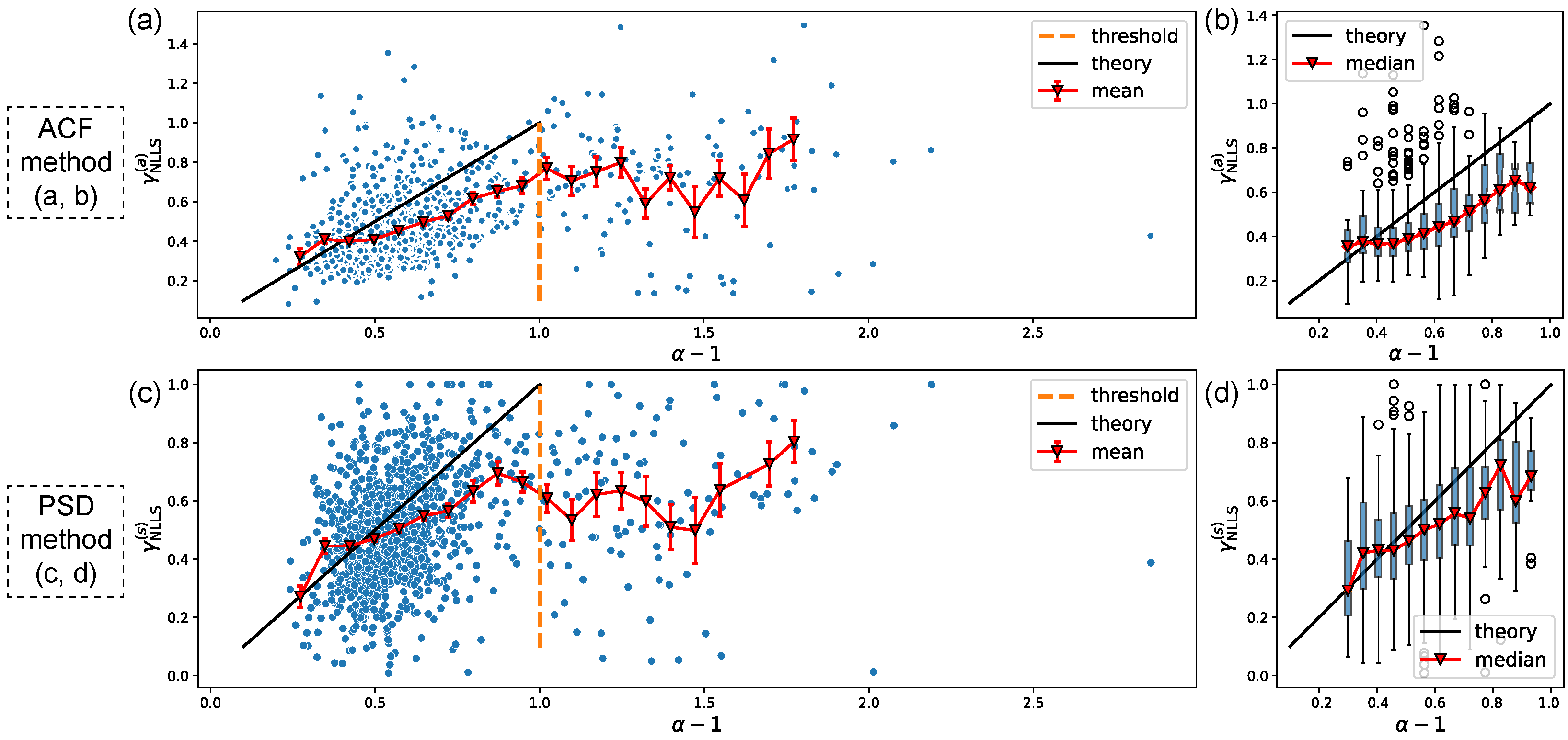"}
		\caption{
			Scatterplot between $\alpha$ and the NLLS estimator $\gamma_{\naive}$ for our dataset based on the ACF method (upper panels, a, b) and the PSD method (lower paners, c, d). 
			(a, c)~The full scatterplot between $\alpha$ and $\gamma_{\naive}$. This figure illustrates that the NLLS estimator is actually biased, particularly for $\alpha>2$.  
			(b, d)~The boxplot between $\alpha-1\in (0,1)$ and $\gamma_{\naive}$. This boxplot shows an approximate linear relation but shows the systematic deviation, as expected by the LMF simulations.
		}
		\label{fig:FullScatterPlot}
	\end{figure}
	
	For our data analysis in the main text, we focused on the scatterplot between $\alpha$ and the naive estimator $\gamma_{\unbiased}$. This is because the NLLS estimator $\gamma_{\naive}$ has a statistical bias, and the unbiased estimator $\gamma_{\unbiased}$ is a better basis for our study. For reference, we show the scatterplot between $\alpha$ and $\gamma_{\naive}$ for our dataset as Figs.~\ref{fig:FullScatterPlot}(a) and (c) for the ACF and PSD methods, respectively. These figures illustrate that the naive estimator $\gamma_{\unbiased}$ is actually biased due to the finite sample size. For reference, we also show the boxplot in as Figs.~\ref{fig:FullScatterPlot}(b) and (d) for the ACF and PSD methods, respectively. As expected, the bias is much more serious for $\alpha>2$. 

\section{Robustness check of our statistical analysis}\label{sec:app:CrossValidation}
	\begin{figure}
		\centering
		\includegraphics[width=170mm]{"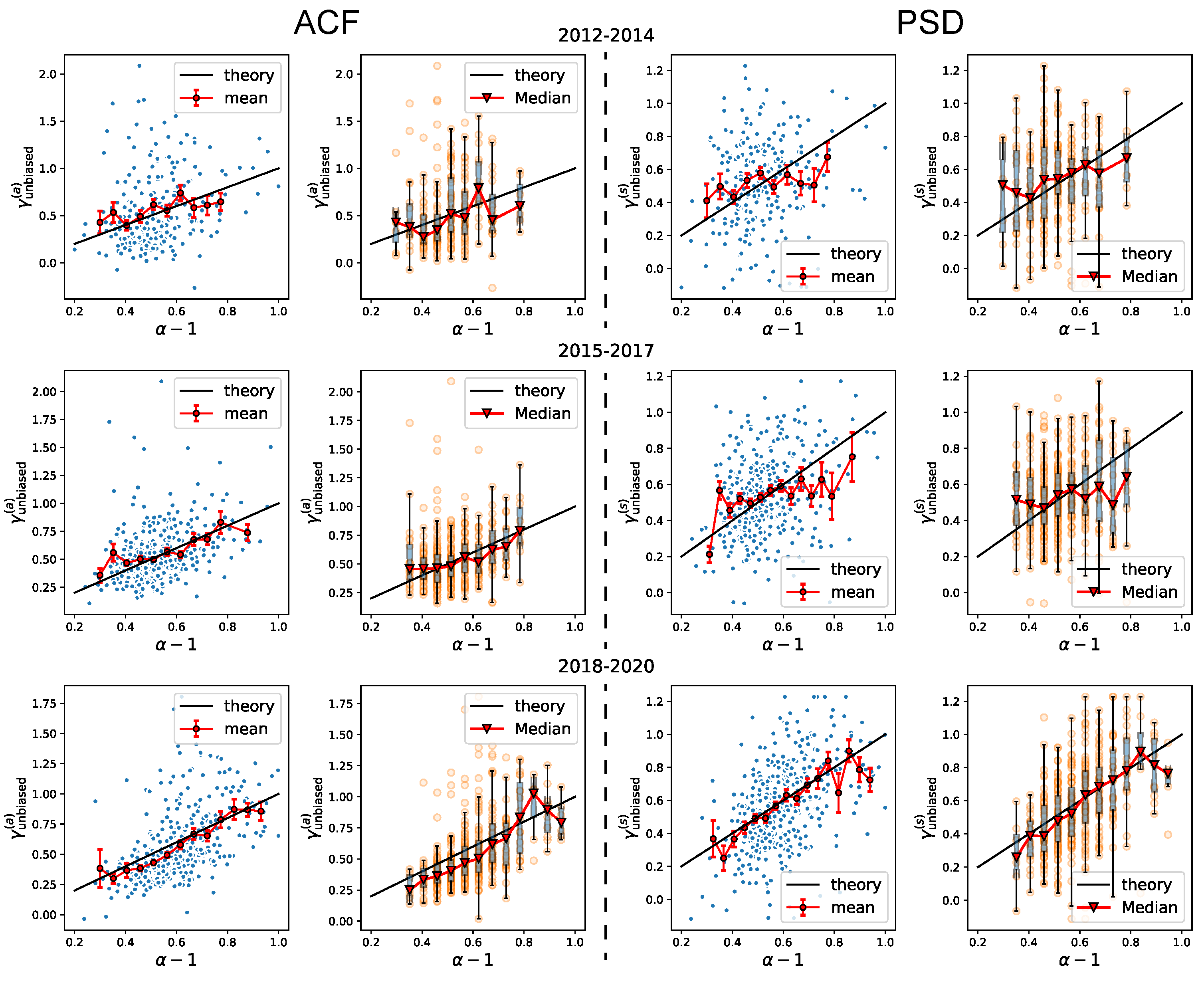"}
		\caption{
			Robustness check of our statistical analysis. Our nine-year dataset was split into three datasets from 2012 to 2014, from 2015 to 2017, and from 2018 to 2020. The unbiased estimator $\gamma_{\unbiased}$ agrees with the theoretical line $\gamma=\alpha-1$ for the three periods. The left (right) panels are based on the ACF (PSD) methods.
		}
		\label{pic:CrossValidation}
	\end{figure}
	In the main text, we tested the validity of the LMF prediction $\gamma=\alpha-1$ by using the nine-years data. On the other hand, it would be more scientifically sound to check its statistical robustness. In this appendix, we examined the temporal robustness of the LMF prediction. 

	For the robustness check, the nine-year dataset was split into three datasets: the datasets (i) from 2012 to 2014, (ii) from 2015 to 2017, and (iii) from 2018 to 2020. We apply the same method as in Sec.~\ref{sec:measure_gamma} to these three independent datasets to test whether the LMF prediction holds for these three periods. The results are summarised in Fig.~\ref{pic:CrossValidation} (see the left (right) panels for the results based on the ACF (PSD) methods). The LMF prediction $\gamma=\alpha-1$ consistently holds for the three independent periods, suggesting the statistical robustness of our results. We have an impression that the goodness of fit improved for the most recent dataset (2018-2020), which might be related to the increasing numbers of transactions, particularly by the STs.

	\section{The LMF unbiased estimator for the total number of the splitting traders}
	\label{sec:app:N_ST_via_LMFprefactor}
		\begin{figure*}
			\centering
			\includegraphics[width=160mm]{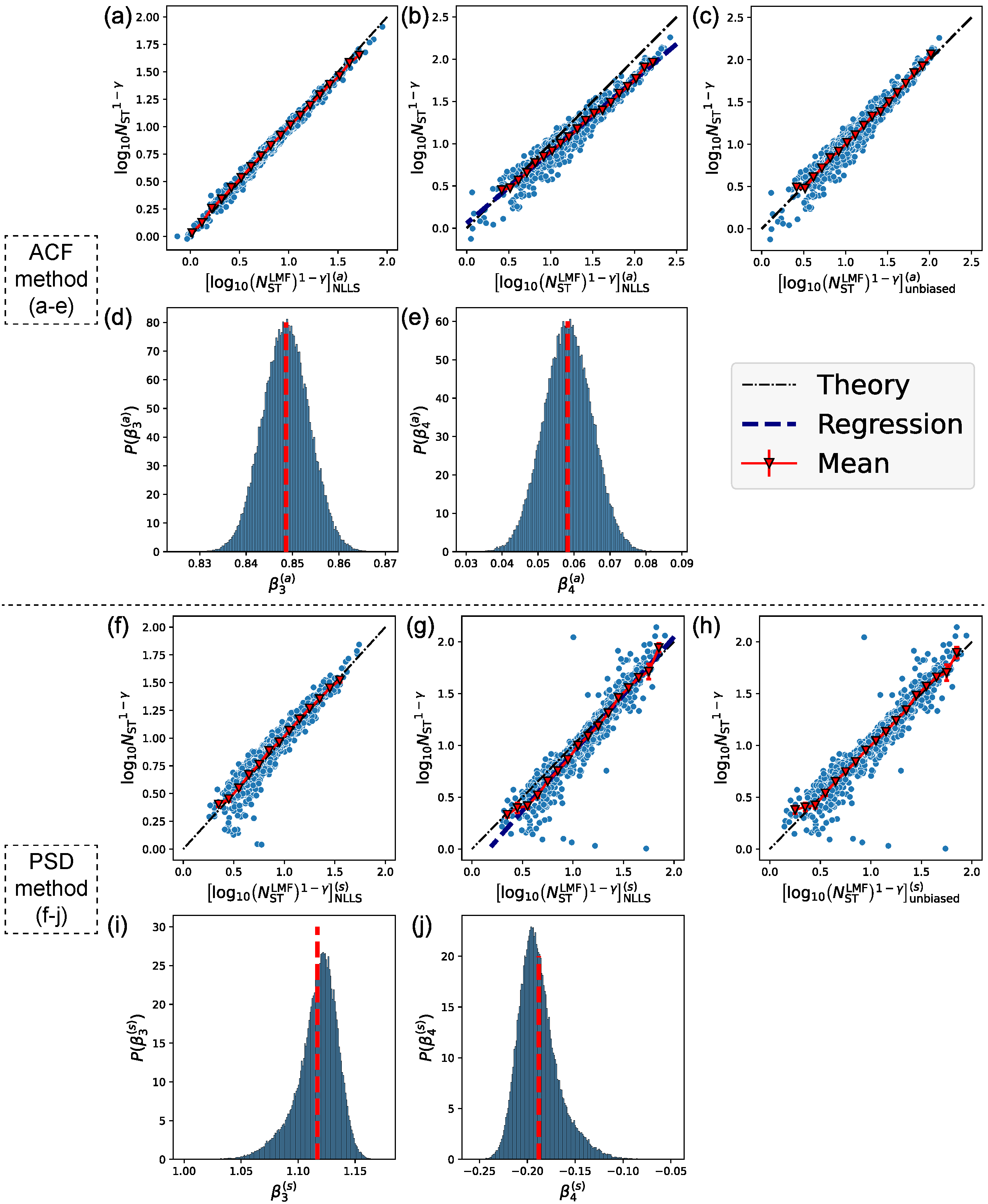}
			\caption{
				Numerical study of the NLLS estimator $[\log_{10}(N_{\ST}^{\LMF})^{1-\gamma}]_{\naive}$ based on the ACF (a-e) and PSD (f-j) methods for the LMF simulations. The superscript $X^{\acf}$ ($X^{\psd}$) signifies that the estimator is based on the ACF (PSD) method. (a)~Consistency check of the ACF-NLLS estimator for $N_{\eps}=10^8$. The parameters $(N_{\ST},\alpha)$ are the same as in our dataset. (b)~The ACF-NLLS estimator is biased for $N_{\eps}\lesssim 10^7$. The parameters $(N_{\ST},\alpha, N_{\eps})$ are the same as in our dataset. (c)~ACF-based unbiased estimator $[\log_{10}(N_{\ST}^{\LMF})^{1-\gamma}]_{\unbiased}$ approximately shows the unbiasedness as expected. (d, e)~Histogram of the regression coefficients $(\beta_3, \beta_4)$ in Eq.~\eqref{eq:app:regression_logN} for the ACF method. (f-j) Corresponding figures for the PSD method. 
			}
			\label{fig:app:c_0-UB}
		\end{figure*}
		This appendix describes the detailed construction method of an LMF unbiased estimator for the total number of the STs $N_{\ST}$. Be they based on the ACF or PSD method, let $\gamma_{\naive}$ and $c_{0,\naive}$ be the NLLS estimators for the ACF power-law exponent and the ACF prefactor as available quantities even from public data. The LMF theory predicts that the total number of traders $N_{\ST}$ is equal to the LMF estimator $N_{\ST}^{\rm LMF}(c_0, \gamma):=[(\gamma+1)c_0]^{1/(\gamma-1)}$. We therefore study the relationship between the true value of $\log_{10}\left(N_{\ST}\right)^{1-\gamma}$ and $\log_{10}\left(N_{\ST}^{\LMF}\right)^{1-\gamma}$. The NLLS estimator is constructed as 
		\begin{equation}
			\left[\log_{10}\left(N_{\ST}^{\LMF}\right)^{1-\gamma}\right]_{\naive}
			:=\log_{10}\left[\frac{1}{(\gamma_{\naive}+1)c_{0,\naive}}\right].
		\end{equation}
	
		\subsection{Consistency for the infinite sample size}
			Let us check the consistency of the NLLS estimator by assuming the LMF model with realistic parameters regarding $(N_{\ST},\alpha)$. We performed the numerical simulations of the LMF model for a sufficiently-large sample size $N_{\eps}=10^8$. We plotted the true values of $\log_{10}\left(N_{\ST}\right)^{1-\gamma_{\naive}}$ for the vertical axis and those of $[\log_{10}(N_{\ST}^{\LMF})^{1-\gamma}]_{\naive}$ for the horizontal axis in Fig.~\ref{fig:app:c_0-UB}(a) (Fig.~\ref{fig:app:c_0-UB}(f)) for the ACF method (the PSD method). The figure shows the agreement between our numerical result and the theory, supporting consistency of the NLLS estimator $[\log_{10}(N_{\ST}^{\LMF})^{1-\gamma}]_{\naive}$. 
		
		\subsection{Bias for the finite sample size}
			On the other hand, if we set realistic parameters regarding $(N_{\ST},\alpha,N_{\eps})$ with $N_{\eps}\lesssim 10^7$, the numerical result in Fig.~\ref{fig:app:c_0-UB}(b) (Fig.~\ref{fig:app:c_0-UB}(g)) shows systematic deviations from the theoretical line. Indeed, let us apply the regression:
			\begin{equation}
				\log_{10}\left(N_{\ST}\right)^{1-\gamma_{\naive}} = \beta_3[\log_{10}(N_{\ST}^{\LMF})^{1-\gamma}]_{\naive} + \beta_4.
				\label{eq:app:regression_logN}
			\end{equation}
			We measured the values of $(\beta_3, \beta_4)$ for 100 times and took their ensemble average as $(\la\beta_3\ra, \la\beta_4\ra)=(0.847,0.058)$ for the ACF method ($(\la \beta_3\ra, \la\beta_4\ra)=(1.117, -0.188)$ for the PSD method). See also Fig.~\ref{fig:app:c_0-UB}(d, e) (Fig.~\ref{fig:app:c_0-UB}(i, j)) for the histogram of $(\beta_3, \beta_4)$ regarding the ACF (PSD) method. This result implies that the NLLS estimator $[\log_{10}(N_{\ST}^{\LMF})^{1-\gamma}]_{\naive}$ is biased due to the finite sample size effect.
		
		\subsection{Construction of unbiased estimators}
			We next construct the unbiased estimators for $\log_{10}\left(N_{\ST}\right)^{1-\gamma_{\naive}}$. The unbiased estimator $[\log_{10}(N_{\ST}^{\LMF})^{1-\gamma}]_{\unbiased}$ is constructed as 
			\begin{equation}
				\left[\log_{10}\left(N_{\ST}^{\LMF}\right)^{1-\gamma}\right]_{\unbiased}
			:= \beta_3 \left[\log_{10}\left(N_{\ST}^{\LMF}\right)^{1-\gamma}\right]_{\naive} + \beta_4,
			\end{equation}
			which shows an approximate unbiasedness by definition
			\begin{equation}
				\left< \left[\log_{10}\left(N_{\ST}^{\LMF}\right)^{1-\gamma}\right]_{\unbiased}\right> 
				\approx \log_{10}\left(N_{\ST}\right)^{1-\gamma_{\naive}}.
			\end{equation}
			See Figs.~\ref{fig:app:c_0-UB}(c) and (h) for the numerical check of the unbiasedness for the LMF simulations. Figure~\ref{fig:emp_estimate_N} is based on this approximate unbiased estimator $[\log_{10}(N_{\ST}^{\LMF})^{1-\gamma}]_{\unbiased}$.

	\section{Nonlinear relative least-squares method}\label{sec:app:RLS_method}
	In our fitting, we use the nonlinear relative least-squares (RLS) method, which is formulated as follows: let us consider the datapoints $\{(\tau_i, y_i)\}_{i=1,\dots, N_{\rm dat}}$ and consider the fitting function $f(\tau|\bm{p})$ with the parameters $\bm{p}:=(p_1,\dots,p_K)$. We fix the optimal parameter $\bm{p^*}$ as 
	\begin{equation}
		\bm{p}^* = \argmin_{\bm{p}} J_{\RLS}(\bm{p}), \>\>\> 
		J_{\RLS}(\bm{p}) := \sum_{i=1}^{N_{\rm dat}} \left(\frac{y_i-f(\tau_i | \bm{p})}{f(\tau_i | \bm{p})}\right)^2,
	\end{equation}
	where $J_{\RLS}(\bm{p})$ is the cost function for the RLS method. 
	Note that the ordinary least-squares (OLS) method is formulated by
	\begin{equation}
		\bm{p}^* = \argmin_{\bm{p}} J_{\OLS}(\bm{p}), \>\>\> 
		J_{\OLS}(\bm{p}) := \sum_{i=1}^{N_{\rm dat}} \left(y_i-f(\tau_i | \bm{p})\right)^2.
	\end{equation}
	Their difference comes from the cost functions between $J_{\RLS}(\bm{p})$ and $J_{\OLS}(\bm{p})$.

	We did not employ the OLS method because, in the OLS method, the tail of the fitting function (i.e., $f(\tau|\bm{p})$ for large $\tau$) much less contributes to the cost function than the head of the fitting function (i.e., $f(\tau|\bm{p})$ for small $\tau$). Since we are interested in the power-law exponent of the tail with $f(\tau|\bm{p})\approx \tau^{-\gamma}$ for large $\tau$, the contribution from the tail should not be underestimated. To clarify this point mathematically, let us rewrite the cost function of the OLS as 
	\begin{equation}
		J_{\OLS}(\bm{p}) := \sum_{i=1}^{N_{\rm dat}} f^2(\tau_i|\bm {p})\left(\frac{y_i-f(\tau_i | \bm{p})}{f(\tau_i | \bm{p})}\right)^2
		= J_{\rm weighted} (\bm{p} \>| \{f^2(\tau|\bm{p})\}_\tau),
	\end{equation}
	where we define the weighted cost function 
	\begin{equation}
		J_{\rm weighted} (\bm{p} \>| \{w(\tau)\}_\tau) := \sum_{i=1}^{N_{\rm dat}} w(\tau_i)\left(\frac{y_i-f(\tau_i | \bm{p})}{f(\tau_i | \bm{p})}\right)^2.
	\end{equation}
	This representation highlights that the tail of the fitting function is much less contributing to the cost function in the OLS method since $w(\tau) \ll w(0)$ for large $\tau$. In contrast, the RLS has a better character than the OLS because the contributions to the cost function are theoretically expected to be the same between the head and the tail. We note that the cost function of the RLS method can be rewritten as 
	\begin{equation}
		J_{\RLS}(\bm{p}) = J_{\rm weighted} (\bm{p} \>| \{1\}_\tau).
	\end{equation}

\section{Smoothing on the logarithmic time axis}\label{sec:app:logarithmic_smoothing}
	Let us consider a smoothing method based on the logarithmic time. For simplicity, let us consider the AFC $C(\tau)$ for the continuous time $\tau \in (0,\infty)$. If the ACF asymptotically obeys the power-law decay $C(\tau)\approx C_0\tau^{-\gamma}$, its log-log plot should be linear as 
	\begin{equation}
		\ln C(\tau) \approx \ln C_0 - \gamma \ln \tau.
	\end{equation}
	Therefore, it is customary to plot the log-log plot of the ACF to confirm the power-law decay. Based on this mathematical fact, we consider a smoothing of the ACF in the logarithmic time: by defining the logarithmic time $x:= \ln \tau$, we introduce a smoothed ACF for a given $\tau$ as
	\begin{equation}
		C_{\smooth}(\tau) := \int_{0}^{\infty} \tilde{w}_\delta(x(\tau); x')C(\tau(x'))dx', \>\>\> \tau(x) := e^{x}, \>\>\> \tau(x') := e^{x'}.  
	\end{equation}
	with the weight function $w_\delta(x;x')$ and the smoothing window size $\delta>0$. By assuming that $\tilde{w}_{\delta}(x;x')$ is uniform on the logarithmic time $x$ as 
	\begin{equation}
		\tilde{w}_\delta(x; x') = \begin{cases}
			\frac{1}{\delta} & (x' \in [x-\delta/2,x+\delta/2)) \\
			0 &  (x' \not \in [x-\delta/2,x+\delta/2))
		\end{cases},
	\end{equation}
	we obtain the ACF formula of the logarithmic smoothing: 
	\begin{equation}
		C_{\smooth}(\tau) = \int_{x(\tau)-\delta/2}^{x(\tau)+\delta/2} \frac{dx'}{\delta}C(\tau(x'))
		= \int_{\tau e^{-\delta/2}}^{\tau e^{+\delta/2}} \frac{d\tau'}{\delta \tau'}C(\tau')
	\end{equation}
	with the variable transformation $\tau' := e^{x'}$. This is equivalent to 
	\begin{subequations}\label{eq:app:smoothedACF_log}
	\begin{equation}
		C_{\smooth}(\tau) = \int_{0}^{\infty} w_{\delta}(\tau;\tau')C(\tau')d\tau', \>\>\> 
		w_{\delta}(\tau;\tau') := \begin{cases}
			\displaystyle
			\frac{1}{\delta \tau'} & (\tau' \in [\tau_{\smooth}^-(\tau), \tau_{\smooth}^+(\tau)))\\
			0 & (\tau' \not \in [\tau_{\smooth}^-(\tau), \tau_{\smooth}^+(\tau)))
		\end{cases}
	\end{equation}
	with 
	\begin{equation}		
		\tau_{\smooth}^-(\tau) := \tau e^{-\delta/2}, \>\>\>
		\tau_{\smooth}^+(\tau) := \tau e^{+\delta/2}.
	\end{equation}
	\end{subequations}
	Thus, Eq.~\eqref{eq:ACF_smooth_log} is the natural extension of the smoothed ACF formula for the discrete time $\tau$.

\end{document}